\documentclass{aa}  

\usepackage{graphicx}
\usepackage{txfonts}
\usepackage{graphicx}	
\usepackage{amsmath}	
\usepackage{amsfonts}
\usepackage{multicol}   
\usepackage{bm}		    
\usepackage{pdflscape}	
\usepackage{acro}
\usepackage[normalem]{ulem}
\usepackage{natbib}
\usepackage{hyperref}
\usepackage[capitalise]{cleveref}
\usepackage{widetext}
\usepackage{xcolor}
\usepackage{placeins}
\usepackage{afterpage}
\usepackage{longtable}

\newcommand{\iap}{CNRS \& Sorbonne Universit\'{e}, Institut d’Astrophysique de Paris (IAP), UMR 7095, 98 bis bd Arago, F-75014 Paris, France}
\newcommand{\cca}{Center for Computational Astrophysics, Flatiron Institute, 162 5th Avenue, New York, NY 10010, USA}
\newcommand{\tsinghua}{Department of Astronomy, Tsinghua University, Beijing 100084, China.}
\newcommand{\jhupha}{Department of Physics and Astronomy, Johns Hopkins University, Baltimore, MD 21218, USA}
\newcommand{\jhuams}{Department of Applied Mathematics and Statistics, Johns Hopkins University, Baltimore, MD 21218, USA}
\newcommand{\icg}{Institute of Cosmology \& Gravitation, University of Portsmouth, Dennis Sciama Building, Portsmouth, PO1 3FX, UK}
\newcommand{\ox}{Astrophysics, University of Oxford, Keble Road, OX1 3RH, UK}

\newcommand{\Mpch}{\ensuremath{h^{-1}\text{Mpc}}}
\newcommand{\classcode}{\textsc{class}}
\newcommand{\camb}{\textsc{camb}}
\newcommand{\colossus}{\textsc{colossus}}
\newcommand{\operon}{\textsc{operon}}

\newcommand{\bacco}{\textsc{bacco}}
\newcommand{\euclidemu}{\textsc{euclidemulator2}}
\newcommand{\cosmopower}{\textsc{cosmopower}}
\newcommand{\quijote}{\textsc{quijote}}
\newcommand{\halofit}{\textsc{halofit}}
\newcommand{\hmcode}{\textsc{hmcode}}
\newcommand{\syren}{\textsc{syren-halofit}}
\newcommand{\syrenplus}{\textsc{syren-new}}
\newcommand{\mksym}[1]{\ifmmode {\rm #1}\else #1\fi}
\DeclareMathOperator{\aq}{aq}
\DeclareMathOperator{\pow}{pow}

\newcommand{\splitatcommas}[1]{%
  \begingroup
  \begingroup\lccode`~=`, \lowercase{\endgroup
    \edef~{\mathchar\the\mathcode`, \penalty0 \noexpand\hspace{0pt plus 1em}}%
  }\mathcode`,="8000 #1%
  \endgroup
}

\begin{document} 

   \title{\syrenplus: Precise formulae for the linear and nonlinear matter power spectra with massive neutrinos and dynamical dark energy}

   \author{
    Ce Sui \thanks{\href{mailto:suic20@mails.tsinghua.edu.cn}{suic20@mails.tsinghua.edu.cn}} \inst{1}
    \and
    Deaglan J. Bartlett \thanks{\href{mailto:deaglan.bartlett@iap.fr}{deaglan.bartlett@iap.fr}} \inst{2}
    \and
    Shivam Pandey \inst{3}
    \and
    {Harry Desmond} \inst{4}
    \and
    {Pedro G. Ferreira} \inst{5}
    \and
    {Benjamin D. Wandelt} \inst{2,3,6,7}
    }

   \institute{
        \tsinghua
        \and
        \iap
        \and
        \jhupha
        \and
        \icg
        \and
        \ox
        \and
        \jhuams
        \and
        \cca
}

   \date{Received XXX; accepted YYY}

 
  \abstract
   {
   Current and future large scale structure surveys aim to constrain the neutrino mass and the equation of state of dark energy. To do this efficiently, rapid yet accurate evaluation of the matter power spectrum in the presence of these effects is essential.
   }
   {
   We aim to construct accurate and interpretable symbolic approximations to the linear and nonlinear matter power spectra as a function of cosmological parameters in extended $\Lambda$CDM models which contain massive neutrinos and non-constant equations of state for dark energy. 
   This constitutes an extension of the \textsc{syren-halofit} emulators to incorporate these two effects, which we call \syrenplus{} (SYmbolic-Regression-ENhanced power spectrum emulator with NEutrinos and $W_0-w_a$).
   We also wish to obtain a simple approximation to the derived parameter $\sigma_8$ as a function of the cosmological parameters for these models.
   }
   {
   We utilise symbolic regression to efficiently search through candidate analytic expressions to approximate the various quantities of interest. Our results for the linear power spectrum are designed to emulate \textsc{class}, whereas for the nonlinear case we aim to match the results of \textsc{euclidemulator2}. We compare our results to existing emulators and $N$-body simulations.
   }
   {
   Our analytic emulators for $\sigma_8$, the linear and nonlinear power spectra achieve root mean squared errors of 0.1\%, 0.3\% and 1.3\%, respectively, across a wide range of cosmological parameters, redshifts and wavenumbers. The error on the nonlinear power spectrum is reduced by approximately a factor of 2 when considering observationally plausible dark energy models and neutrino masses. We verify that emulator-related discrepancies are 
   subdominant compared to observational errors and other modelling uncertainties when computing shear power spectra for LSST-like surveys. Our expressions have similar accuracy to existing (numerical) emulators, but are at least an order of magnitude faster, both on a CPU and GPU.
   }
   {
   Our work greatly improves the accuracy, speed and range of applicability of current symbolic approximations to the linear and nonlinear matter power spectra. These now cover the same range of cosmological models as many numerical emulators with similar accuracy, but are much faster and more interpretable. 
   We provide publicly available code for all symbolic approximations found.
   }


    \titlerunning{\syrenplus}
   \maketitle
%

\section{Introduction}
\label{sec:Introduction}

Despite the phenomenal successes of the standard models of cosmology ($\Lambda$CDM) and particle physics, we do not yet fully understand the nature of many of their constituents. In particular, it is known that the Universe is expanding and this expansion is accelerating, yet we do not know what the properties of the component of the Universe (dark energy) which drives this are.
The current cosmological paradigm states that this is due to a cosmological constant, which has an equation of state which does not vary in time and is equal to -1, yet recent observations hint at deviations from these assumptions \citep{DESI_2024}.
On smaller scales, we have, to date, detected three flavour eigenstates of neutrinos and oscillation experiments demonstrate that these can be mixed
\citep[e.g.][]{Becker-Szendy_1992,Fukuda_1998a,Fukuda_1998b,Ahmed_2004}. This implies that neutrinos must have non-zero mass. Although we can measure the differences between the neutrino masses with these techniques, the masses of the individual flavours remain unknown. We do not even know the order (hierarchy) of their masses, namely if it is ``normal'' ($m_1 \leq m_2 \leq m_3$, where $m_i$ are the mass eigenstates and the smallest difference between the square of the masses is between $m_1$ and $m_2$) or ``inverted'' ($m_3 \leq m_1 \leq m_2$).

One of the main aims of current and future large scale structure surveys such as \textit{Euclid} \citep{Euclid_2011}, LSST \citep{LSST_2009}, DESI \citep{DESI_2016} and the Nancy Grace Roman Space Telescope \citep{WFIRST_2019} is to shed light on these fundamental questions.
In particular, by studying their effects on cosmological scales \citep[see, e.g.][]{Lesgourgues_2006,Weinberg_2013}, these experiments aim to constrain the equation of state of dark energy and the total mass of neutrinos. 

Central to current large scale structure analyses is the matter power spectrum; the Fourier transform of the two-point correlation function.
By comparing the theoretical and observed power spectra, one can constrain cosmological parameters through techniques such as Markov Chain Monte Carlo. 
The matter power spectrum also serves as an input to the calculation of many other cosmological observables such as the shear power spectrum for weak lensing analyses or the galaxy power spectrum for clustering studies.
In the absence of any accelerated modelling techniques, to compute the theoretical prediction down to very nonlinear scales accurately, one would have to run a new $N$-body simulation each time one updated the cosmological parameters. Or, even if one was only interested in relatively linear scales, one would still need to solve a complicated set of differential equations \citep{Lewis_2000,Blas_2011,Hahn_2023} to make this prediction.
Given that these inferences require one to sample cosmological parameters many thousands of times, a significant amount of work has been dedicated to bypassing these requirements to significantly increase the speed of cosmological analyses.

Many of the techniques exploited to expedite these calculations involve the construction of ``emulators'': typically advanced numerical machine learning based approaches (e.g. Gaussian Processes or Neural Networks) or interpolators designed to be used as surrogate models which directly output the power spectrum given a set of cosmological parameters.
Some examples of these emulators are
\textsc{Aemulus} \citep{Zhai_2019},
\bacco{} \citep{Angulo_2021,Arico_2021,Zennaro_2023},
\textsc{cobra} \citep{Bakx_2024},
\textsc{CosmicEmu} \citep{Lawrence_2017},
\cosmopower{} \citep{SpurioMancini_2022}.
\textsc{Dark Emulator} \citep{Nishimichi_2019},
\textsc{emuPK} \citep{Mootoovaloo_2022},
\textsc{euclidemulator1} \citep{Knabenhans_2019},
\euclidemu{} \citep{Knabenhans_2021},
\textsc{FofrFittingFunction} \citep{Winther_2019},
\textsc{FrankenEmu} \citep{Heitmann_2009,Heitmann_2014},
\textsc{NGenHalofit} \citep{Smith_2019},
and \textsc{pico} \citep{Pico2,Pico1}.
Although accurate to the percent level and considerably faster than the calculations which they are designed to replace, these have disadvantages particularly regarding interpretability.
It is not clear how a given feature of the final prediction is sensitive to the input parameters, and it challenging to ensure that the emulated quantities have the correct asymptotic behaviour in well-known limits.

As an alternative, many symbolic approximations to the linear and nonlinear matter power spectra have been proposed. 
Arguably the most widely-used are the approximations by \citet{Eisenstein_1998,Eisenstein_1999}, which give linear matter power spectra which are accurate to a few percent.
These superseded the less accurate approximation by \citet{Bardeen_1986} (BBKS) which did not include the effects of baryonic acoustic oscillations (BAOs).
Analytic approximations to the nonlinear power spectrum have also been constructed, which rely on the halo model formalism \citep{Ma_2000,Seljak_2000,Cooray_2002}.
Assuming that matter in the Universe is bound in dark matter halos, one can either make this prediction by performing integrals of several quantities such as the halo density profile and mass function, as in the \hmcode{} approach \citep{Mead_2015,Mead_2016,Mead_2021}, or one can adopt the \halofit{} method \citep{Smith_2003,Bird_2012,Takahashi_2012}, where one directly predicts $P(k)$ using an analytic function of the linear power spectrum and cosmological parameters.

These traditional symbolic approaches are not sufficiently accurate for modern applications, where percent-level predictions are required \citep{Taylor_2018}.
Hence, symbolic regression (SR; for a recent review see \citealt{Kronberger_2024}) has recently been employed as a method to improve these models by automatically and rapidly searching through candidate symbolic approximations to the matter power spectrum.
Building off the approximation of \citet{Eisenstein_1998}, \citet{Bartlett_2024_linear} were able to obtain a symbolic emulator for the linear matter power spectrum of $\Lambda$CDM cosmologies which is accurate to 0.1\% for $k = 9\times10^{-3} - 9 \, h{\rm \, Mpc^{-1}}$ and across a wide range of cosmological parameters.
Similar work was performed by \citet{Bayron-Orjuela-Quintana_2023} who obtained an expression for the linear matter power spectrum in modified gravity models which has a typical accuracy of around 2\% away from the BAOs, where the error is larger.
The same authors extended their fit to include these features \citep{Orjuela-Quintana_2024} for universes with modified gravity and massive neutrinos, although this fit only achieves an accuracy of 3\% in the power spectrum\footnote{The authors report an error of 1.5\% on the transfer function at $k\sim 0.1 \, h {\rm \, Mpc^{-1}}$ for this cosmology. The power spectrum is proportional to the square of this quantity, hence the error on the power spectrum is twice as large as this value.} near the first acoustic peak for the Planck 2018 cosmology \citep{Planck_VI_2018}, so does not meet the desired percent-level accuracy.
To improve nonlinear predictions for the power spectrum, \citet{Bartlett_2024_syren} introduced the \syren{} model: a SR-enhanced version of \halofit{} which is accurate to 1\% for $\Lambda$CDM. Not only does this model have comparable accuracy to numerical emulators, but it was found to be between 36 and 950 times faster than alternative methods.

Given the desire to constrain deviations from $\Lambda$CDM in large scale structure studies, in this work we extend the work of \citet{Bartlett_2024_linear,Bartlett_2024_syren} to obtain analytic approximations for the linear and nonlinear power spectra for cosmologies with non-zero neutrinos mass and a time-dependent equation of state for dark energy.
We call this model \syrenplus{} (SYmbolic-Regression-ENhanced power spectrum emulator with NEutrinos and $W_0-w_a$).
We base our linear power spectrum emulator on the approximation of \citet{Bartlett_2024_linear}, where we fit for the residuals between the predictions for the extended cosmology and that of $\Lambda$CDM.
Unlike \citet{Bartlett_2024_syren}, for the nonlinear power spetcrum, we do not fit corrections to existing \halofit{} formulae but allow SR to find a new, more accurate fitting formula without relying on previous expressions.
As in \citet{Bartlett_2024_linear}, we also obtain a symbolic approximation to $\sigma_8$ (a derived parameter) as a function of the other cosmological parameters.
All our emulators achieve percent-level accuracy, making them comparably accurate to numerical approaches and to our previous approximations \citep{Bartlett_2024_linear,Bartlett_2024_syren}.
These expressions are not only interpretable, but have added advantages compared to their numerical counterparts; namely, they are faster to evaluate, more portable (they are easy to implement in the user's favourite programming language) and potentially have better longevity (they do not become outdated when the underlying packages become deprecated since they only use standard mathematical operators). 

The paper is organised as follows. In \cref{sec:Theory} we briefly introduce the matter power spectrum and discuss the cosmological models we are concerned with in this work. We then provide a short summary of SR in \cref{sec:symbolic_regression}. Our emulators are constructed in the following three sections, with \cref{sec:sigma8_emulator} detailing the approximation for $\sigma_8$, and \cref{sec:linear_emulator,sec:nonlinear_emulator} describing the emulators for the linear and nonlinear power spectra, respectively.
Our results are compared to pre-existing approximations and emulators in terms of speed and accuracy in \cref{sec:Performance}, where we also study the correlation of our emulation errors with cosmological parameters and the impact on weak lensing studies when such errors are propagated through to shear power spectra.
We conclude in \cref{sec:Conclusion}.
The main results of this paper are given in \cref{eq:sigma8_result,eq:plin_R_result,eq:plin_S_result,eq:pk_nl_fit,eq:pnl_offset,eq:pnl_fit_final}.
Throughout this paper ``$\log$'' denotes the natural logarithm, and base-10 logarithms are denoted by ``$\log_{10}$''.

\section{Theoretical Background}
\label{sec:Theory}

Our goal is to develop an efficient, differentiable, and, if feasible, interpretable emulator for the power spectrum of matter fluctuations in the Universe, denoted as $P(k, a, \bm{\theta})$. Here, $k$ represents the wavenumber, $\bm{\theta}$ stands for the cosmological parameters, and $a$ is the scale factor ($a = 1 / (1+z)$ for redshift $z$).

Separating the time-dependent matter density $\rho(\bm{x}, a, \bm{\theta})$ in the Universe into a spatially constant background density, $\bar{\rho}(a, \bm{\theta})$, and a density contrast, $\delta(\bm{x},a, \bm{\theta})$, such that $\rho(\bm{x},a, \bm{\theta}) = \bar{\rho}(a, \bm{\theta})[1 + \delta(\bm{x}, a, \bm{\theta})]$, the power spectrum is defined as follows. If $\tilde{\delta}(\bm{k},a, \bm{\theta})$ denotes the Fourier Transform of $\delta(\bm{x},a, \bm{\theta})$, and the matter distribution is statistically homogeneous and isotropic, it follows that
\begin{equation}
    \label{eq:pk_definition}
    (2 \pi)^3 P(k, a, \bm{\theta}) \delta^{\rm D} \left( \bm{k} - \bm{k}^\prime \right)\equiv\langle \tilde{\delta}(\bm{k}, a, \bm{\theta}) \tilde{\delta}^\ast (\bm{k}^\prime, a, \bm{\theta}) \rangle,
\end{equation}
where $\langle\cdots\rangle$ denotes an ensemble average and $\delta^{\rm D}$ is the Dirac delta function.

In this work we consider two versions of the matter power spectrum. For the ``linear'' power spectrum, which we denote as $P_{\rm lin}(k, a, \bm{\theta})$, \cref{eq:pk_definition} is evaluated using $\tilde{\delta}(\bm{k}, a, \bm{\theta})$ as computed using linear perturbation theory.
The second variant uses the fully nonlinear value of $\tilde{\delta}(\bm{k}, a, \bm{\theta})$ as is thus named the ``nonlinear'' power spectrum, and denoted by $P_{\rm nl}(k, a, \bm{\theta})$ in this paper.

Although much of the focus of this paper will be on obtaining symbolic approximations to $P(k,a,\bm{\theta})$, we will also find an emulator for $\sigma_8$ as a function of the other cosmological parameters. This is defined using
\begin{equation}
    \label{eq:sigmaR}
    \sigma_R^2 \equiv \int {\rm d}k \, \frac{k^2}{2 \pi^2} P_{\rm lin} (k, a=1, \bm{\theta}) \left| W(k,R) \right|^2,
\end{equation}
where the Fourier transfer of the top-hat filter is 
\begin{equation}
    W(k, R) = \frac{3}{(kR)^3} \left( \sin (k R) - kR \cos (k R) \right),
\end{equation}
and $\sigma_8$ is $\sigma_R$ for $R = 8 \Mpch$.
Performing such an integral can be time consuming, especially if one is, for example, post-processing a Markov Chain Monte Carlo analysis for which this is a derived parameter.
Alternatively, if one required $A_{\rm s}$ for a given $\sigma_8$, the usual approach involves computing the linear matter power spectrum using a Boltzmann code with an initial estimate of $A_{\rm s}$, then performing the integral in \cref{eq:sigmaR} to determine $\sigma_8$. For a desired $\sigma_8$ value, $\sigma_8^\prime$, one then uses $A_{\rm s}^\prime = (\sigma_8^\prime/\sigma_8)^2 A_{\rm s}$. Again, this can be time consuming.

This work considers an extended version of the $\Lambda$CDM model, described by the parameters given in \cref{tab:cosmo_par_prior}.
This model is an extension of $\Lambda$CDM in two ways.
First, we study a Universe containing three massive neutrinos of equal mass, such that the sum of their massess is given by $m_\nu$, which is measured in ${\rm eV}$ throughout this work.
Second, instead of assuming that dark energy is explained through a cosmological constant, $\Lambda$, we allow it to have a time-dependent equation of state. We adopt the commonly used Chevallier-Polarski-Linder (CPL) parameterisation \citep{Chevallier_2000,Linder_2002}, such that
\begin{equation}
    \label{eq:dark_energy_eos}
    w(a) = w_0 + w_a (1 - a),
\end{equation}
where $\Lambda$CDM corresponds to $w_0=-1$ and $w_a=0$.
Given the small dependence of the power spectrum on the reionisation optical depth parameter, $\tau$, we do not vary this in this work.

The ranges of the cosmological parameters given in \cref{tab:cosmo_par_prior}  are almost identical to those given in Table 2 of \citet{Knabenhans_2021}, except that we use a slightly smaller range of $w_a$ ($w_a < 0.5$ instead of $w_a < 0.7$). Although \citet{Knabenhans_2021} state they allow these larger values of $w_a$, in fact they only train \euclidemu{} on the same parameter range as \cref{tab:cosmo_par_prior}, hence our emulators are trained on exactly the same range of cosmological models.
Values of $w_a$ larger than 0.5 give very different predictions than the rest of parameter space, since they allow models with $w_{0}+w_{a}=0$, and thus it is extremely challenging to obtain a single emulator covering all models if these values are included.
Given that these models are not viable given current data \citep{DESI_2024}, this slight reduction of parameter space does not significantly weaken the applicability of our results.

In our previous papers \citep{Bartlett_2024_linear,Bartlett_2024_syren}, we computed the linear power spectrum and $\sigma_8$ using \camb{} \citep{Lewis_2000}. However, extending to $w_0-w_a$ is problematic since this involves some models for which $w < -1$, which is not permitted in \camb.  
We therefore use \classcode{} \citep{Blas_2011} to compute the power spectrum variables of interest in this work since it can deal with these situations.

\begin{table*}
    \caption{Cosmological parameters, redshifts and wavenumbers used for analytic emulators.}
    \centering
    \begin{tabular}{c|l|c|c}
        Parameter & Description & Minimum & Maximum \\
        \hline\hline
         $10^9 \, A_{\rm s}$ & Spectral amplitude. & 1.7 & 2.5\\
         $\Omega_{\rm m}$ & Total matter (cold dark matter + baryons + neutrinos) density parameter. & 0.24 & 0.40 \\
         $\Omega_{\rm b}$ & Baryonic density parameter. & 0.04 & 0.06 \\
         $h$ & Dimensionless Hubble parameter ($H_0 = h \times 100 {\rm \, km \, s^{-1} \, Mpc^{-1}}$). & 0.61 & 0.73 \\
         $n_{\rm s}$ & Spectral index. & 0.92 & 1.00 \\
         $w_0$ & Time independent part of the dark energy equation of state (\cref{eq:dark_energy_eos}). & -1.3 & -0.7 \\
         $w_a$ & Time dependent part of the dark energy equation of state (\cref{eq:dark_energy_eos}). &  -0.7 & 0.5 \\
         $m_\nu \ / \ {\rm eV}$ & Sum of neutrino massess (assuming three degenerate species). & 0.00 & 0.15 \\
         $z$ & Redshift. & 0.0 & 3.0 \\
         $k \ / \ h{\, \rm Mpc^{-1}}$ & Wavenumber. & $9 \times 10^{-3}$ & $9$ \\
    \end{tabular}
    \tablefoot{All parameters are sampled independently and uniformly in the range between the minimum and maximum values given.}
    \label{tab:cosmo_par_prior}
\end{table*}

\section{Symbolic regression}
\label{sec:symbolic_regression}

To derive analytical approximations for the linear and nonlinear power spectra and for $\sigma_8$, we utilise the supervised machine learning method of symbolic regression (SR) \citep{Kronberger_2024}.
In particular, we employ the widely-used genetic programming approach \citep{turing,David, haupt}.
Genetic programming involves the evolution of ``computer programs'', in our context, mathematical expressions represented as expression trees. 
Based on the principle of natural selection, at a given iteration, the least effective equations (as determined by a fitness metric) are eliminated, while new equations are generated through the combination of sub-expressions from the current population (crossover) or by randomly modifying subtrees within an expression (mutation). This process continues over numerous generations, gradually evolving the population of equations to produce increasingly accurate analytical expressions.
We choose to use the SR package \operon{}\footnote{\url{https://github.com/heal-research/operon}} \citep{Burlacu_2020} due to its speed, memory efficiency, and strong benchmark performance \citep{LaCava_2021,Burlacu_2023}.
This code has been used successfully in many cosmological and astrophysical studies \citep{Bartlett_2024_linear,Bartlett_2024_syren,Russeil_2024,AbdusSalam_2024}.

SR is typically viewed as a Pareto-optimisation problem, where one attempts to find accurate yet simple descriptions of the data.
In our case, accuracy is determined by the root mean squared error between the predicted and target variable.
To characterise simplicity, we use the ``length'' of the model. In \operon, equations are encoded as trees, with each terminal node (e.g., $k$ or a cosmological parameter) being accompanied by a scaling parameter, which is then optimised~\citep{Kommenda_2020} using the Levenberg–Marquardt algorithm \citep{Levenberg_1944,Marquardt_1963}. The ``length'' of an expression is defined to be the total number of nodes in the tree excluding the scaling terms.

During non-dominated sorting (NSGA2), \operon{} utilises $\epsilon$-dominance \citep{Laumanns_2002}, where $\epsilon$ is defined such that objective values (the accuracy and simplicity metrics) within $\epsilon$ of each other are considered equal. This parameter impacts the number of duplicate equations in the population, promoting convergence to a well-distributed approximation of the global Pareto front: the set of solutions that cannot be improved in accuracy without increasing complexity. We experimented with different values for this parameter to determine settings that yield accurate yet compact models for our emulators, and we give the used values in the relevant sections for each one.

To select the best-fitting model, we first generate the Pareto front of candidate expressions. We then focus on models with a loss below a predetermined threshold (ensuring adequate accuracy for our purposes) and those with similar losses on training and validation sets (indicating minimal over-fitting).
Although model selection can be automated \citep[see, e.g.][]{Bartlett_2023,Bartlett_2022,cranmer2020discovering}, we opt to manually inspect the most accurate solution for each model length. This allows us to qualitatively select a function that is both sufficiently compact for interpretability and accurate enough for our needs.

\section{Analytic emulator for \texorpdfstring{$\sigma_8$}{sigma8}}
\label{sec:sigma8_emulator}

We begin by examining the most basic emulator one might need for quantities related to the power spectrum: an emulator for $\sigma_8$ as a function of other cosmological parameters, or equivalently, an emulator for $A_{\rm s}$ given $\sigma_8$ and the remaining cosmological parameters. 
Our goal is to expedite \cref{eq:sigmaR} with a symbolic emulator.
This is an extension of the emulator found by \citet{Bartlett_2024_linear}, who obtained a corresponding expression for $\Lambda$CDM which was accurate to 0.1\%.

To find an emulator for our extended cosmology, we constructed a Latin hypercube (LH) sampling of 2000 sets of cosmological parameters, using uniform priors within the ranges specified in \cref{tab:cosmo_par_prior} (although we do not vary $z$, since we define $\sigma_8$ at redshift zero). 
We created a second LH sampling of 2000 points for validation purposes. For these parameters, we computed $\sigma_8$ using \classcode{} \citep{Blas_2011} and attempted to learn this mapping using a root mean squared error loss function with \operon.

Given that the primordial power spectrum is proportional to the parameter $A_{\rm s}$, through \cref{eq:sigmaR} one knows that $\sigma_8 \propto \sqrt{A_{\rm s}}$. As such, following \citet{Bartlett_2024_linear}, instead of directly fitting for $\sigma_8$, we choose our target variable to be $\sigma_8 / \sqrt{10^{9} A_{\rm s}}$, which we fit for as a function of the other cosmological parameters. Using $10^{9}A_{\rm s}$ instead of $A_{\rm s}$ enables the target variable to be $\mathcal{O}(1)$.
Since we wish to find relatively compact expressions, we set the maximum model length to be 100. We found that a value of $\epsilon = 10^{-3}$ was appropriate for this problem and allowed expressions to be formed of the operators $+$, $-$, $\times$, $\div$, $\sqrt{\cdot}$, $\pow$ and $\log$.

\begin{figure}
    \centering
    \includegraphics[width=\columnwidth]{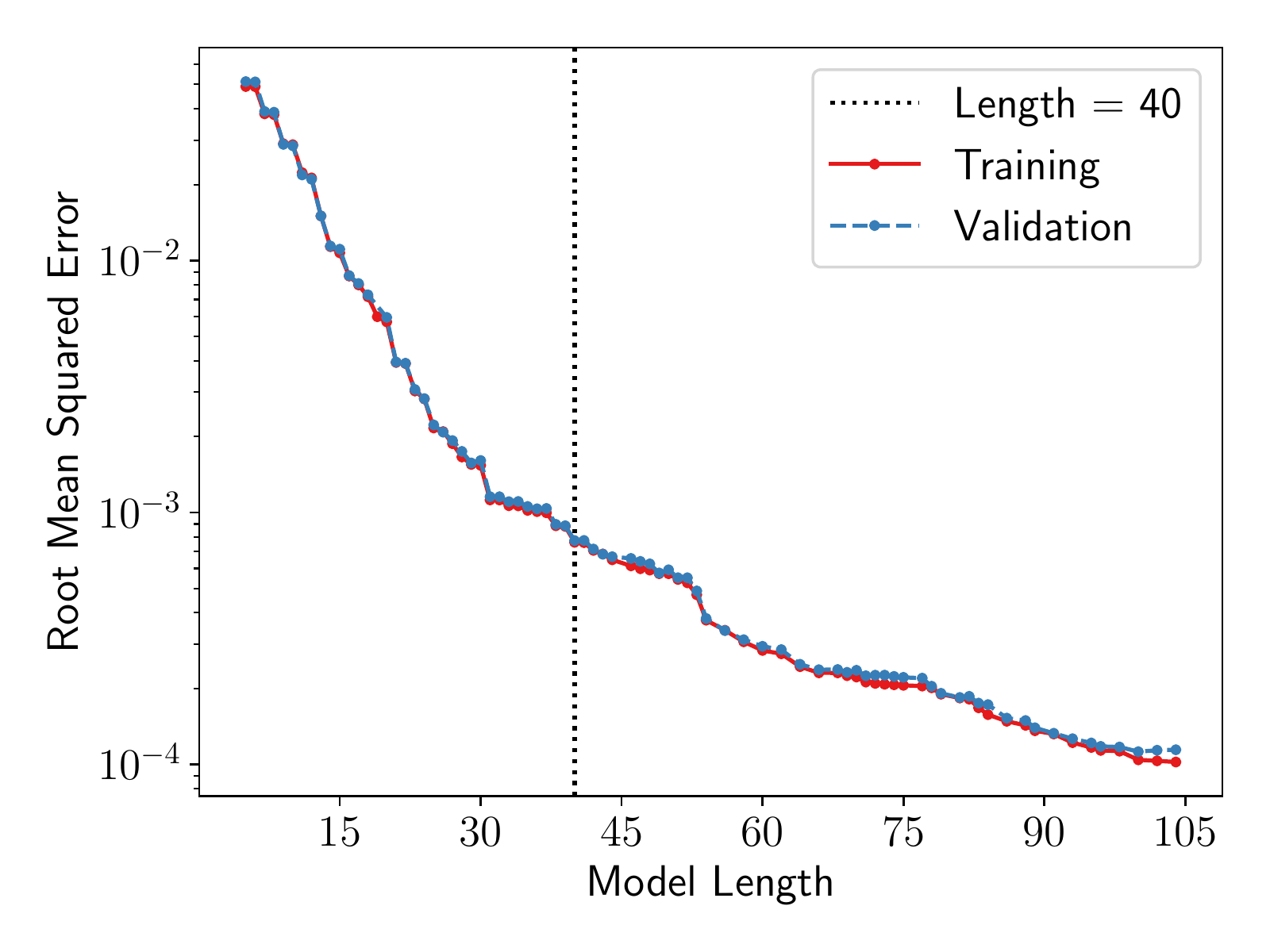}
    \caption{Pareto front of solutions found by \operon{} for $\sigma_8/\sqrt{10^9 A_{\rm s}}$ as a function of cosmological parameters. We plot the curves for our training and validation sets separately, and indicate the chosen model (\cref{eq:sigma8_result}) with a dotted vertical line.}
    \label{fig:sigma8_pareto}
\end{figure}

After searching for 1 hour with 56 cores, \operon{} returned the Pareto front given in \cref{fig:sigma8_pareto}. We see that, for all model lengths, the training and validation losses are approximately equal, indicating that the models are not over-fitting. This is a particular advantage of SR, where the regularity on the solution space imposed by the requirement to use standard mathematical expressions makes over-fitting unlikely. 

We observe that the error on the prediction decreases rapidly as one initially increases the model length, but this improvement becomes slower after a model length of approximately 30, where the root mean squared error is already at 0.1\%. Given that we do not expect to require models which are much more accurate than this, to balance accuracy and simplicity, we choose to select a model from this region, although we note that more accurate models are achievable if one further increased the model length. 

As such, after inspecting these equations, we choose the model with length 40 as our fiducial result. This is given by
\begin{equation}
\label{eq:sigma8_result}
\begin{split}
    \frac{\sigma_8}{\sqrt{10^9 A_{\rm s}}} & \approx 
    c_{0}\left( c_{1}\Omega_{\rm b} + c_{2} \Omega_{\rm m} + \log{\left(c_{3} w_{0} + \log{\left(c_{4} w_{0} + c_{5} w_{a} \right)} \right)}\right) 
    \\&\times \left(c_{6} \Omega_{\rm m} + c_{7} m_{\rm \nu} + c_{8} n_{\rm s} - \log{\left( c_{9} \Omega_{\rm m} + c_{10} w_{a} \right)}\right)\\
    &\times \left( c_{11} \Omega_{\rm b} + c_{12} \Omega_{\rm m} -  n_{\rm s}\right) 
    \left(c_{13} \Omega_{\rm m} + c_{14} h + c_{15} m_{\rm \nu} + n_{\rm s}\right), 
\end{split}
\end{equation}
with the parameters $\bm{c}$ given in \cref{tab:sigma8_params}.
This expression achieves a root mean squared error on 0.1\% on both the training and test sets.
We give the most accurate expression we find in \cref{sec:most_accrurate_sigma8}, which has a root mean squared error of 0.02\% on the validation set. We choose not to make this our fiducial model due since it is a long expression, although we encourage the user to use whichever model is preferable given their target accuracy.

\begin{table}[]
    \caption{Best-fit parameters for the $\sigma_8$ emulator given in \cref{eq:sigma8_result}.}
    \centering
    \begin{tabular}{c|l|c|l|c|l}
    Param. & Value & Param. & Value & Param.  & Value \\
    \hline\hline
    $c_{0}$ & 0.0187 & $c_{6}$ & 1.3057 & $c_{12}$ & -11.1463\\
    $c_{1}$ & -2.4891 & $c_{7}$ & 0.0885 & $c_{13}$ & -1.5433 \\
    $c_{2}$ & 12.9495 & $c_{8}$ & 0.1471 & $c_{14}$ & -7.0578\\
    $c_{3}$ & -0.7527 & $c_{9}$ & 3.4982 & $c_{15}$ & 2.0564\\
    $c_{4}$ & -2.3685 & $c_{10}$ & -0.006 & & \\
    $c_{5}$ & -1.5062 & $c_{11}$ & 19.2779 & & \\
    \end{tabular}
    \label{tab:sigma8_params}
\end{table}

\begin{figure}
    \centering
    \includegraphics[width=\columnwidth]{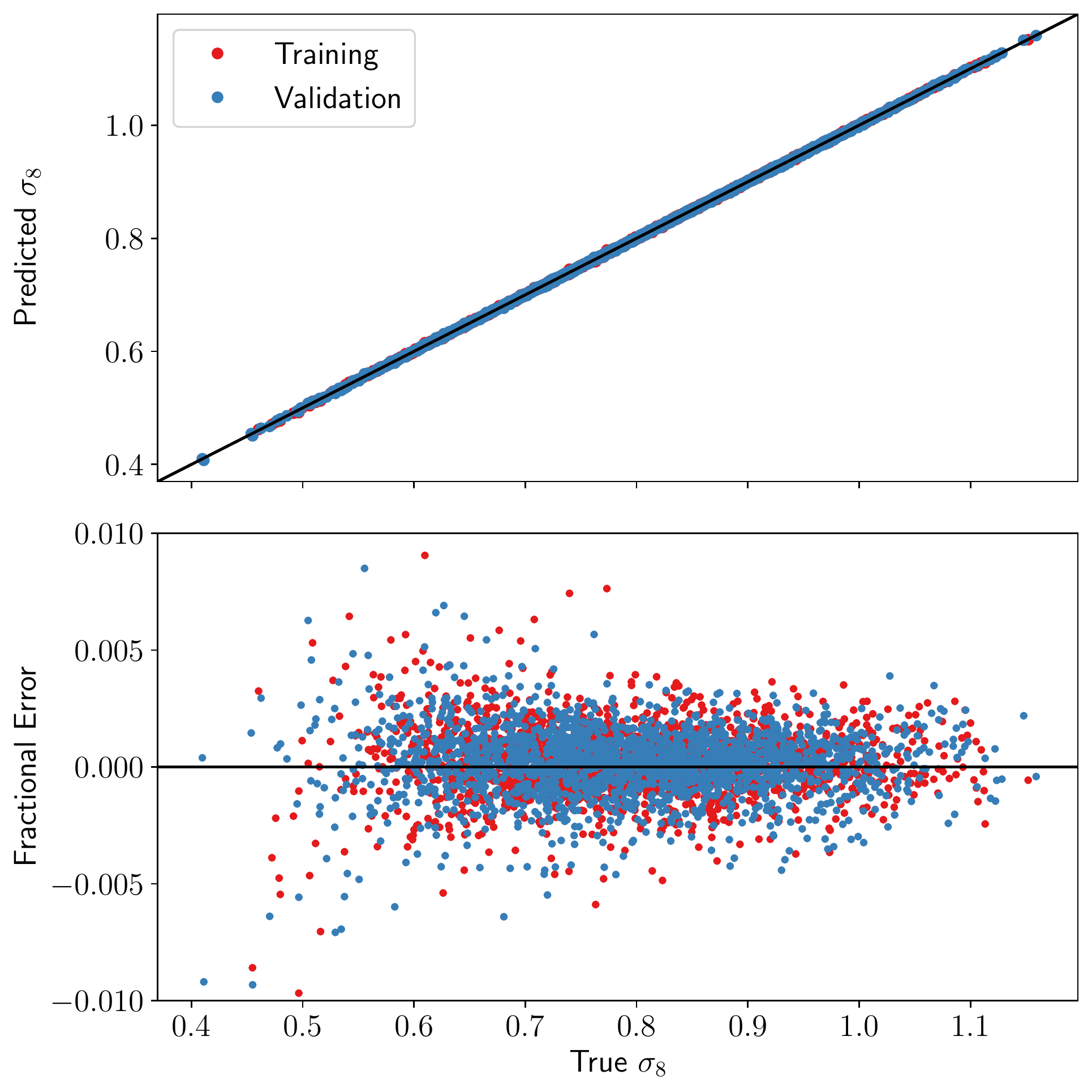}
    \caption{
    Predicted values (upper) and fractional errors (lower) for the $\sigma_8$ emulator (\cref{eq:sigma8_result}) across the training and validation sets. The root mean squared error is 0.1\%, with a maximum error of 1\% for very small $\sigma_8$ values.
    }
    \label{fig:sigma8_prediction}
\end{figure}

To further investigate the performance of this emulator, in \cref{fig:sigma8_prediction} we plot the predicted $\sigma_8$ for each point on the training and validation LH as a function of the true value, and we also give the fractional error.
We see that the fractional error on $\sigma_8$ is approximately constant as the true values are varied, although for very small values of $\sigma_8$ the error slightly increases to a maximum deviation of 1\% at $\sigma_8 \approx 0.5$. These small values are incompatible with current observations \citep{Planck_VI_2018}, and thus this small increase in error is unimportant for actual applications of this result.

\begin{figure}
    \centering
    \includegraphics[width=\columnwidth]{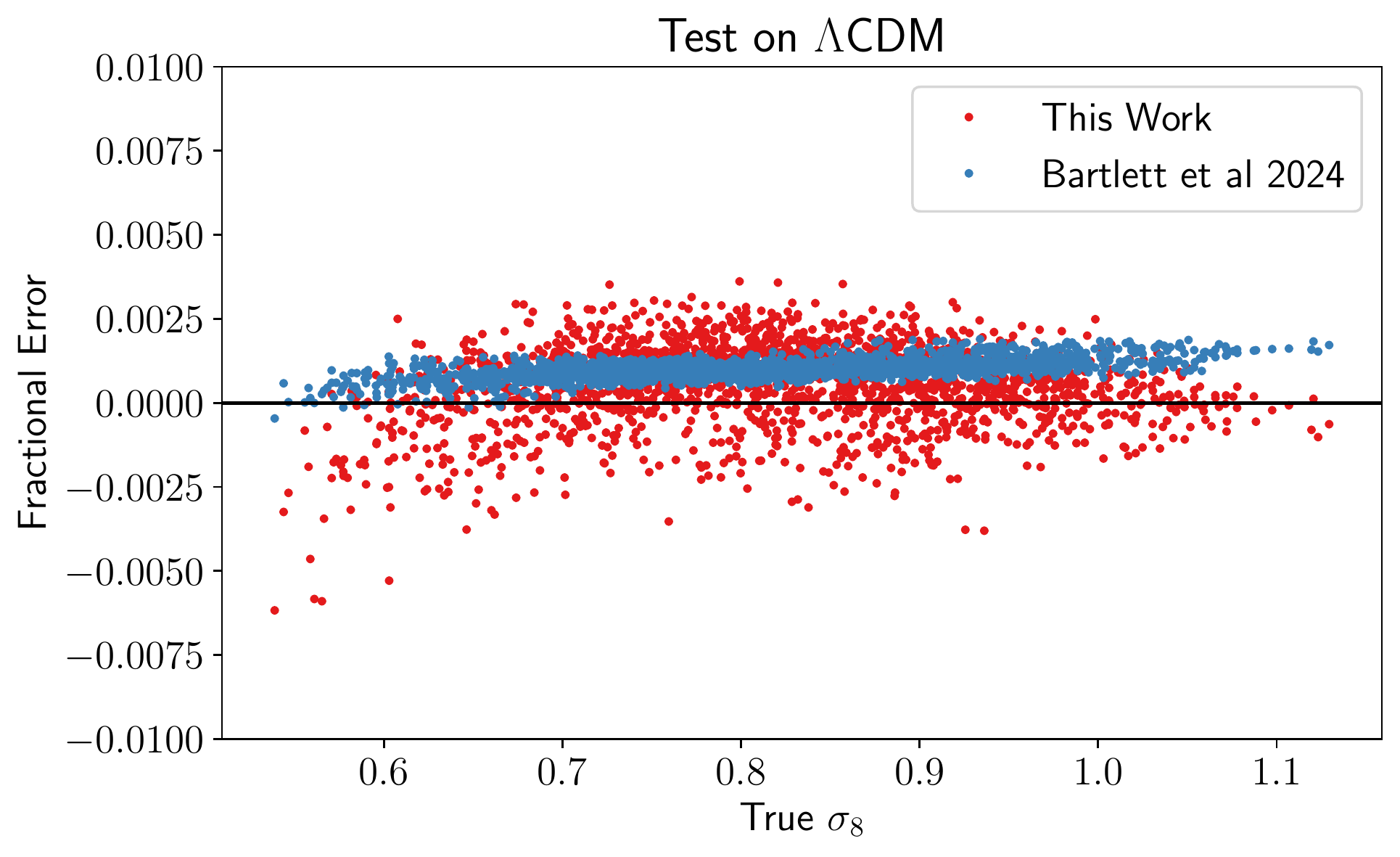}
    \caption{
    Fractional error on the $\sigma_8$ prediction for a test set which is restricted to $\Lambda$CDM models only. We compare the results of the emulator given here to that of \citet{Bartlett_2024_linear}. Our emulator has slightly weaker performance given its extended validity, but the error is almost always within 0.5\%.
    }
    \label{fig:sigma8_lcdm}
\end{figure}

Although we have produced this emulator for an extended cosmology, it is interesting to verify the performance on this emulator in the $\Lambda$CDM subspace to ensure that it can be applied to this situation. To do this, we generate a new LH of 2000 cosmologies within the $\Lambda$CDM portion of \cref{tab:cosmo_par_prior} and compute the corresponding $\sigma_8$ values with \classcode. These are then compared against the predictions of \cref{eq:sigma8_result} in \cref{fig:sigma8_lcdm}, where we also show the results of using the emulator proposed in \citet{Bartlett_2024_linear}.

As one may expect, given that the emulator of \citet{Bartlett_2024_linear} is specifically optimised for $\Lambda$CDM, whereas our method applies to a wider range of cosmologies, we obtain slightly larger emulation errors.
We obtain almost identical root mean squared errors of approximately 0.1\%, but the distribution of errors has slightly larger tails for this work compared to \citet{Bartlett_2024_linear}; our predictions remain almost always within 0.5\%, whereas the \citet{Bartlett_2024_linear} emulator has a maximum error of approximately 0.25\%. 
We observe that \citet{Bartlett_2024_linear} systematically overestimates the results. This is due to our use of \classcode{} instead of \camb{} as the ground truth in this paper and since we included neutrino species in \classcode{} even when $m_{\nu} = 0$, which introduces slight deviations from the $\Lambda$CDM calculations used in \citet{Bartlett_2024_linear} to obtain their result.

\section{Analytic emulator for the linear power spectrum}
\label{sec:linear_emulator}

We now consider the task of obtaining a symbolic approximation for the linear matter power spectrum. 
In the spirit of \citet{Bartlett_2024_linear}, rather than attempt to learn $P_{\rm lin} \left( k, a, \bm{\theta} \right)$ in its entirety, we use some pre-existing approximations for the linear matter power spectrum, so that we only have to learn small corrections to these physics-informed formulae.

To incorporate previous approximations, we find that it is useful to write the linear matter power spectrum as
\begin{equation}
    \label{eq:plin_terms}
    P_{\rm lin} \left( k, a, \bm{\theta} \right)
    =
    D_{\rm approx} \left( k, a, \bm{\theta} \right)^2
    R \left( a, \bm{\theta} \right)
    S \left( k, \bm{\theta} \right)
    P_{\rm \Lambda CDM} \left( k, \bm{\theta}_{\rm \Lambda CDM} \right),
\end{equation}
where $\bm{\theta}$ contains all cosmological parameters, whereas $\bm{\theta}_{\rm \Lambda CDM} = \{ A_{\rm s}, \Omega_{\rm m}, \Omega_{\rm b}, h, n_{\rm s} \}$ are the subset of parameters describing $\Lambda$CDM.
We will utilise existing results for two of these terms: 
$D_{\rm approx}$ (an approximation for the growth factor) and $P_{\rm \Lambda CDM}$ (an approximation for the $\Lambda$CDM linear power spectrum).
We note that $P_{\rm \Lambda CDM}$ gives the shape of the linear matter power spectrum, which is equal to redshift-zero power spectrum up to some normalisation, which the other terms provide.

We therefore have to learn two terms which give corrections to this:
\begin{enumerate}
    \item $R \left( a, \bm{\theta} \right)$: This corresponds to (the square of) the correction to the linear growth factor approximation, $D_{\rm approx}$.
    \item $S \left( k, \bm{\theta} \right)$: This is the correction to the redshift-zero linear power spectrum when moving from $\Lambda$CDM to the more general cosmological models.
\end{enumerate}
These terms are defined such that $R \left( a=1, \bm{\theta} \right) = 1 \, \forall \,\bm{\theta}$. 
Although one could make $R$ depend on $k$, we found that the symbolic expressions we obtained when allowing for this dependence did not include $k$, and therefore we do not include this argument of the function.

We choose to fit for these corrections to $\Lambda$CDM rather than repeat the full fitting procedure of \citet{Bartlett_2024_linear} given that we expect the shape of the baryonic acoustic oscillations to be independent of the neutrino mass and dark energy equation of state. Given that we already have a precise approximation to these in $\Lambda$CDM, it is simpler to fit the broad-band corrections to the shape of the power spectrum which should vary slowly in $k$ and relatively smoothly in time.

In this section, we begin by describing the approximations we use from the literature to give 
 $D_{\rm approx}$ (\cref{sec:Dapprox}) and
$P_{\rm \Lambda CDM}$ (\cref{sec:pk_lcdm}).
We then fit the correction terms to this in \cref{sec:linear_sr}.

\subsection{Approximation for the growth factor}
\label{sec:Dapprox}

To obtain the time-evolution of the linear matter power spectrum, we wish to have an approximation which we can correct with symbolic regression. Fortunately such an approximation already exists. We begin by defining the equality redshift as
\begin{equation}
    z_{\rm eq} = 2.5 \times 10^4 \Omega_{\rm m} h^2 \Theta_{2.7}^{-4}.
\end{equation}
Then, defining the time-dependent dark energy and dark matter density parameters as
\begin{align}
    \Omega_{\rm m}(a) &= \frac{\Omega_{\rm m} a^{-3}}{\Omega_{\rm m} a^{-3} + \left( 1 - \Omega_{\rm m} \right) a^{-3(1 + w_0+w_a)} \exp(- w_a (1-a))} , \\
    \Omega_{\rm \Lambda}(a) &= \frac{\left( 1 - \Omega_{\rm m} \right) a^{-3(1 + w_0+w_a)} \exp(- w_a (1-a)) }{\Omega_{\rm m} a^{-3} + \left( 1 - \Omega_{\rm m} \right) a^{-3(1 + w_0+w_a)} \exp(- w_a (1-a))},
\end{align}
we use the 
following approximation for the growth factor in the absence of neutrinos \citep{Lahav_1991,Carroll_1992,Eisenstein_1998}
\begin{equation}
    \begin{split}
        D_1 (a, \bm{\theta}) = 
        \frac{1 + z_{\rm eq}}{1 + z} & \frac{5 \Omega_{\rm m}(a)}{2} \Bigg(\Omega_{\rm m}(a)^{4/7}  \\
        & - \Omega_{\rm \Lambda}(a) + \left( 1 + \frac{\Omega_{\rm m}(a)}{2} \right)  \left( 1 + \frac{\Omega_{\rm \Lambda}(a)}{70} \right) \Bigg)^{-1}.
    \end{split}
\end{equation}

If $m_{\nu} \neq 0$, then we must correct this to describe the suppression by neutrinos. We will always assume that we have $N_{\rm \nu} = 3$ neutrinos when the neutrino mass is non-zero. One can then define the density parameters of neutrinos and cold dark matter as \citep{Kolb_1990}
\begin{equation}
    \Omega_{\rm \nu} = \frac{m_{\rm \nu}}{93.14 h^2}, \quad
    \Omega_{\rm c} = \Omega_{\rm m} - \Omega_{\rm b} - \Omega_{\rm \nu},
\end{equation}
and then the fractional parameters for each species as
\begin{equation}
    f_{\rm c} \equiv \frac{\Omega_{\rm c}}{\Omega_{\rm m}},
    \quad
    f_{\rm b} \equiv \frac{\Omega_{\rm b}}{\Omega_{\rm m}},
    \quad
    f_{\rm \nu} \equiv \frac{\Omega_{\rm \nu}}{\Omega_{\rm m}},
    \quad
    f_{\rm cb} \equiv f_{\rm c} + f_{\rm b}.
\end{equation}
We then utilise the approximations for the effects of neutrinos on the growth factor of \citet{Bond_1980,Hu_1998}, which are
\begin{align}
    p_{\rm cb} &= \frac{1}{4} \left( 5 - \sqrt{1 + 24 f_{\rm cb}} \right), \\
    \tilde{q} &= \frac{k h \Theta_{2.7}^2}{\Omega_{\rm m}h^2}, \\
    y_{\rm fs} & = 17.2 f_{\rm \nu} \left(1 + \frac{0.488}{f_{\rm \nu}^{7/6}} \right)  \left(\frac{N_{\rm \nu} \tilde{q}}{f_{\rm \nu}}\right)^2, \\
    D_{\rm cb\nu} (k, a, \bm{\theta}) &= \left(f_{\rm cb}^{0.7/p_{\rm cb}} + \left( \frac{D_1 (a, \bm{\theta})}{1 + y_{\rm fs}}\right)^{0.7} \right)^{p_{\rm cb}/0.7} D_1(a, \bm{\theta})^{1 - p_{\rm cb}}.
\end{align}

An approximation to the growth factor can then be written as
\begin{equation}
    \label{eq:D_approx}
    D_{\rm approx} (k, a, \bm{\theta}) = D_{\rm cb\nu} (k, a, \bm{\theta}) \left( 1 + z_{\rm eq} \right)^{-1},
\end{equation}
where we choose a normalisation such that $D_{\rm approx} \to a$ as $a \to 0$.

\subsection{\texorpdfstring{$\Lambda$}{Lambda}CDM power spectrum}
\label{sec:pk_lcdm}

To give an approximate template for the matter power spectrum at redshift zero, we will use the results of \citet{Eisenstein_1998,Bartlett_2024_linear}. To do this, we define $\omega_{\rm m} \equiv \Omega_{\rm m} h^2$ and $\omega_{\rm b} \equiv \Omega_{\rm b} h^2$, and define the following variables
\begin{align}
    s &\equiv \frac{44.5 \log \left( \frac{9.83}{\omega_{\rm m}} \right)}{\sqrt{1.0 + 10.0 \omega_b^{3/4}}}, \\
    \alpha &\equiv 1.0 - 0.328 \log \left(431.0 \omega_{\rm m} \right) \omega_{\rm b} + 0.38 \log \left (22.3 \omega_m\right) \omega_{\rm b}^2 ,\\
    \Gamma &\equiv \Omega_{\rm m} h \left(\alpha + \frac{1.0 -\alpha}{1.0 + (0.43 k h s)^4} \right), \\
    \Theta_{2.7} &= \frac{T_{\rm CMB}}{2.7 {\rm \, K}}, \\
    q &\equiv k \Theta_{2.7} \Gamma^{-1}\\
    C_0 &\equiv 14.2 + \frac{731.0}{1 + 62.5q} \\
    L_0 &\equiv \log \left( 2e + 1.8 q \right),
\end{align}
where $k$ is measured in $h\,{\rm Mpc^{-1}}$ and $T_{\rm CMB}$ is the temperature of the CMB (which we assume to be $2.7255 {\rm \, K}$ throughout \citep{Mather_1999}). Given this, one can write the \citeauthor{Eisenstein_1998} ``no-wiggles'' power spectrum at $z=0$ as
\begin{equation}
    \label{eq:EH_nw}
    P_{\rm EH} (k, \bm{\theta}_{\rm \Lambda CDM}) =
    \frac{2 \pi^2}{k^3} A_{\rm s} \left(\frac{k}{k_{\rm pivot}} \right)^{n_{\rm s} - 1}
    \left(\frac{2 k^2 c^2}{5 \Omega_{\rm m}} 
    \frac{L_0}{L_0 + C_0 q^2}\right)^2,
\end{equation}
where $k_{\rm pivot}=0.05 \,h {\rm  Mpc^{-1}}$.
This describes the overall envelope of the matter power spectrum correctly, but does not contain baryonic acoustic oscillations.

To approximate the redshift-zero power spectrum in $\Lambda$CDM, we define
\begin{equation}
    \label{eq:Pk_residual_definition}
    P_{\rm \Lambda CDM} (k, \bm{\theta}_{\rm \Lambda CDM}) \equiv P_{\rm EH}(k, \bm{\theta}_{\rm \Lambda CDM}) F(k, \bm{\theta}_{\rm \Lambda CDM}),
\end{equation}
where we use the approximation of \citet{Bartlett_2024_linear} to evaluate $F(k, \bm{\theta})$:
\begin{equation}
    \label{eq:lcdm_linear_F}
    \begin{split}
        & \log F =  b_0 h - b_{1} + \left( \frac{b_2 \Omega_{\rm b}}{\sqrt{h^2 + b_3}} \right)^{b_4 \Omega_{\rm m}} \\
       & \times \Biggl[
            \frac{b_5 k - \Omega_{\rm b}}{\sqrt{b_6 + (\Omega_{\rm b} - b_7 k)^2}} b_8 (b_9 k)^{-b_{10} k} 
            \cos \left( b_{11} \Omega_{\rm m} - \frac{b_{12} k}{\sqrt{b_{13} + \Omega_{\rm b}^2}} \right) \\ 
            & - b_{14} \left( \frac{b_{15} k}{\sqrt{1 + b_{16} k^2}} - \Omega_{\rm m} \right) \cos \left( \frac{b_{17} h}{\sqrt{1 + b_{18} k^2}} \right)
        \Biggl] \\
        & + b_{19} (b_{20} \Omega_{\rm m} + b_{21} h - \log(b_{22} k) + (b_{23} k)^{- b_{24} k}) \cos \left( \frac{b_{25}}{\sqrt{1 + b_{26} k^2}} \right) \\
        & + (b_{27} k)^{-b_{28} k} \left( b_{29} k - \frac{b_{30} \log(b_{31} k)}{\sqrt{b_{32} + (\Omega_{\rm m} - b_{33} h)^2}} \right) \\
        & \times \cos \left(  b_{34} \Omega_{\rm m} - \frac{b_{35} k}{\sqrt{b_{36} + \Omega_{\rm b}^2}} \right),
\end{split}
\end{equation}
where the parameters $\bm{b}$ are given in Table 2 of \citet{Bartlett_2024_linear}.
This approximation gives a linear power spectrum which has a root mean squared fractional error of 0.2\% across the $k$ values and $\Lambda$CDM portion of the parameter range considered in this work \citep{Bartlett_2024_linear} when multiplied by the appropriate linear growth factor.

\subsection{SR corrections to the linear power spectrum}
\label{sec:linear_sr}

The approximations outlined in \cref{sec:pk_lcdm,sec:Dapprox} yield reasonable approximations to the matter power spectrum, but do not have sub-percent precision. Even when restricted to the $\Lambda$CDM subspace, since $D_{\rm approx}$ is only an approximation, we find that errors on $P_{\rm lin}(k)$ can still be up to a few percent. Therefore, in this section we use SR to correct for these errors.

\subsubsection{Corrections to the growth factor}

To obtain a correction to the growth factor, $R(a, \bm{\theta})$, we generate two LHs of 2000 cosmologies from \cref{tab:cosmo_par_prior}, one for training and one for validation. 
We used 200 logarithmically spaced $k$ values in the range $9 \times 10^{-3} - 9 \, h{\, \rm Mpc^{-1}}$ and compute $R$, as defined in \cref{eq:plin_terms}, and using the results of \cref{sec:pk_lcdm,sec:Dapprox}.
We find the mean over these $k$ values, and this is then fitted as a function of cosmological parameters and scale factor using \operon, where we search for expressions containing the operators $+$, $-$, $\times$, $\div$, $\sqrt{\cdot}$ and $\log$. 
We allow a maximum model length of 100 and set $\epsilon = 10^{-3}$.

\begin{figure*}
  \centering
   \includegraphics[width=0.48\textwidth]{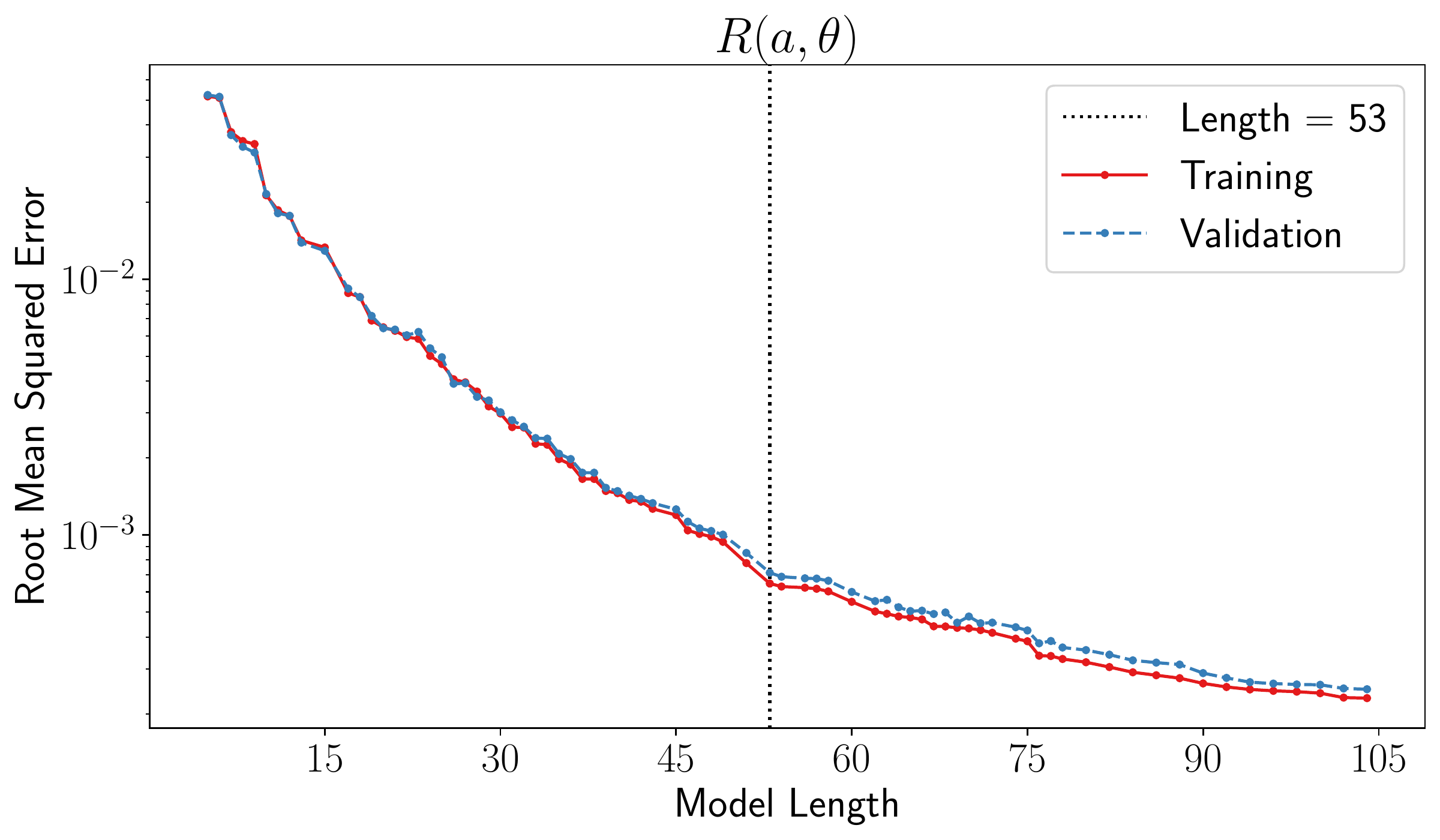}
   \includegraphics[width=0.48\textwidth]{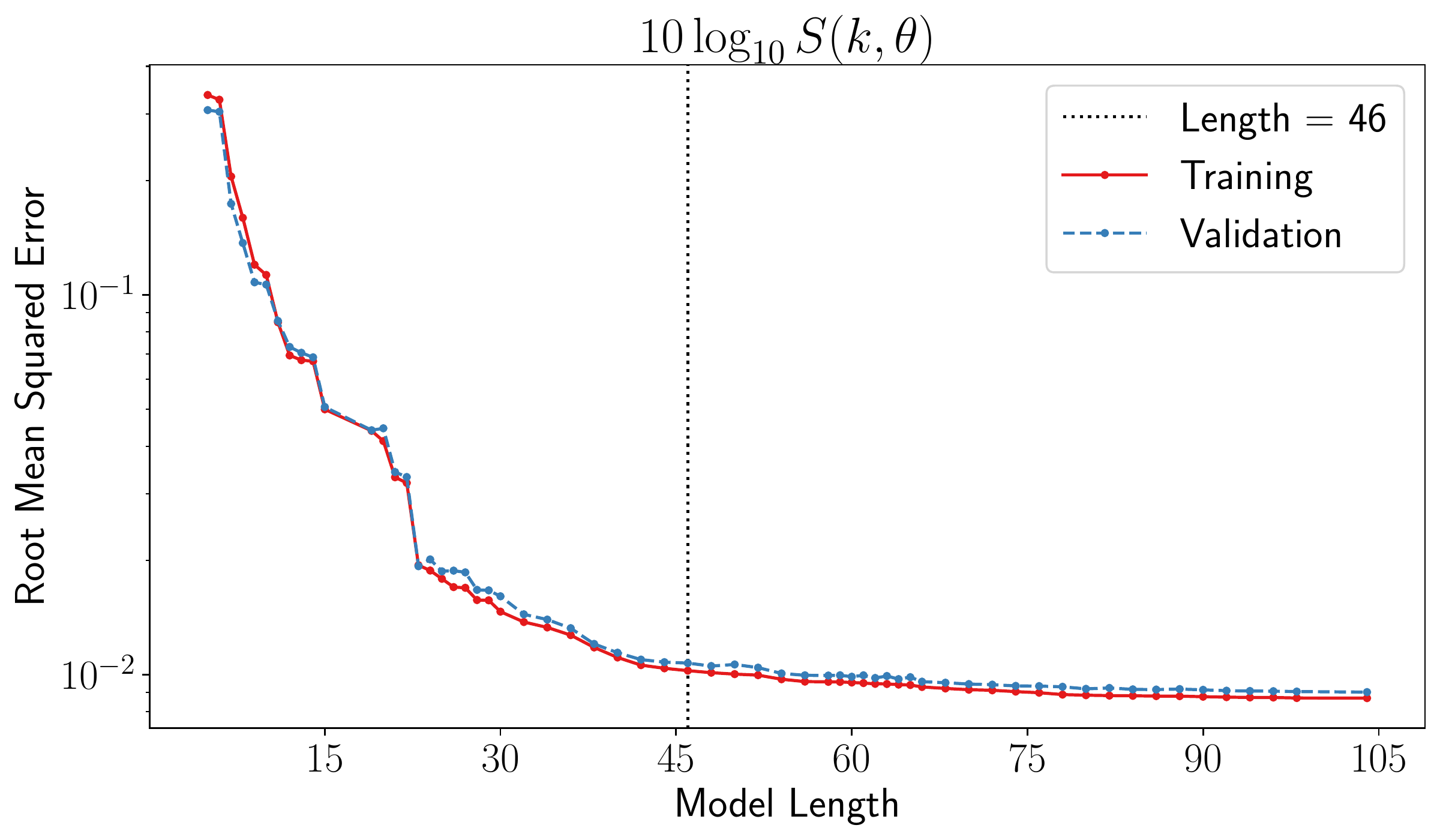}
  \caption{Pareto front of solutions for the correction to the growth factor, $R(a,\bm{\theta})$, (left), and the correction to the redshift-zero matter power spectrum, $10 \log_{10} S(k,\bm{\theta})$, (right) as found by \operon. The chosen models are indicated by the vertical lines, and we plot the root mean squared errors on the predictions for the training and validation sets separately. 
  }
  \label{fig:plin_pareto}  
\end{figure*}

After running for 1 hour on 56 cores, we obtain the Pareto front plotted in \cref{fig:plin_pareto}. As with our $\sigma_8$ emulator, we find very similar training and validation losses for all model lengths. The improvement in accuracy as one makes the model more complex slightly plateaus at a length of $\sim 50$, at which point the root mean squared error is below 0.1\%. We therefore choose a model for $R$ from this region of the Pareto front, and, after inspection, choose the model of length 53.

Although we defined $R(a=1, \bm{\theta}) = 1 \, \forall \bm{\theta}$, the SR result does not return this exactly due to the finite precision of the parameters and the fact that this is only inferred from the training data, rather than strictly enforced during the search. As such, we find that some parameters which are approximately equal must be equated and that some parameters must be rounded in order to require this condition exactly.
After performing such manipulations, we arrive at the following approximation for $R(a,\bm{\theta})$
\begin{equation}
\label{eq:plin_R_result}
\begin{split}
    R & \left( a, \bm{\theta} \right) \approx 
    1 + (1-a) \biggl[ d_0 \\ 
    & - \frac{1}{d_1 a + d_2 + \left( d_3 \Omega_{\rm m} + d_4 a \right) \log{\left(d_5 w_0 + d_6 w_a \right)}} 
    \\&- \frac{d_7 \Omega_{\rm m}  + d_8 a  + \log{\left(d_9 w_0 + d_{10} w_a \right)}}{d_{11} a + d_{12} + d_{13}\left(d_{14} \Omega_{\rm m} + d_{15} a  - 1\right) \left(d_{16} w_0 + d_{17} w_a + 1\right)} 
    \biggr],
\end{split}
\end{equation}
where the parameters $\bm{d}$ are given in \cref{tab:plin_R_params}. It is trivial to see that this equals unity at $a=1$ for any set of cosmological parameters.

\begin{table}[]
    \caption{Best-fit parameters for the correction to the growth factor approximation given in \cref{eq:plin_R_result}.}
    \centering
    \begin{tabular}{c|l|c|l|c|l}
    Param. & Value & Param. & Value & Param.  & Value \\
    \hline\hline
    $d_{0}$ & 0.8545 & $d_{6}$ & -0.4136 & $d_{12}$ & 5.8014\\
    $d_{1}$ & 0.394 & $d_{7}$ & 1.4769 & $d_{13}$ & 6.7085\\
    $d_{2}$ & 0.7294 & $d_{8}$ & -0.5959 & $d_{14}$ & 0.3445\\
    $d_{3}$ & 0.5347 & $d_{9}$ & -0.4553 & $d_{15}$ & 1.2498\\
    $d_{4}$ & -0.4662 & $d_{10}$ & -0.0799 & $d_{16}$ & 0.3756\\
    $d_{5}$ & -4.6669 & $d_{11}$ & -5.8311 & $d_{17}$ & 0.2136\\
    \end{tabular}
    \label{tab:plin_R_params}
\end{table}

This expression is interesting since, although we allowed \operon{} to obtain a function of any combination of cosmological parameters, we find that only $\Omega_{\rm m}$, $w_0$ and $w_a$ appear in \cref{eq:plin_R_result}. Given that these are the parameters which govern the expansion history (alongside $h$), this suggests that this equation is correcting for the mistakes made in modelling the background expansion when using $D_{\rm approx}$.
Although $m_{\rm \nu}$ do not appear in this equation, it does not mean that there is is no time-dependence in the effects of neutrino suppression on $P_{\rm lin}(k,a,\bm{\theta})$, but just that $D_{\rm approx}$ captures this effect sufficiently well for sub-percent level predictions.

\subsubsection{Corrections to the present-day linear power spectrum}
\label{sec:plin_S}

The next step is to obtain a correction, $S( k, \bm{\theta})$, to the redshift-zero power spectrum. This correction depends on $k$, therefore the training set size is expanded by the number of $k$ values,
making it challenging to use a large training size. We found that LHs of 200 cosmologies with 200 logarithmically spaced $k$ values in the range $9 \times 10^{-3} - 9 \, h{\, \rm Mpc^{-1}}$ is sufficient for training an accurate $S( k, \bm{\theta})$.  We then compute $S$, as defined in \cref{eq:plin_terms}, and using the results of \cref{sec:pk_lcdm,sec:Dapprox}.
We choose our target variable to be $10 \log_{10}S(k,\bm{\theta})$ so that (i) the emulated $S(k,\bm{\theta})$ will always be positive, (ii) a root mean squared error on the target corresponds to minimising the fractional error on $S(k,\bm{\theta})$, and (iii) we introduce the factor of 10 to make the target closer to unity.
This is then fitted as a function of cosmological parameters and $k$ values using \operon, where we search for expressions containing the operators $+$, $-$, $\times$, $\div$, $\sqrt{\cdot}$, $\log$, $\cos$, $\pow$ and analytic quotient operators ($\aq (x,y)\equiv x/\sqrt{1+y^2}$). We allow a maximum model length of 100 and set $\epsilon = 10^{-3}$. 

After running for 6 hour on 56 cores, we obtain the Pareto front plotted in \cref{fig:plin_pareto}.
We see that the Pareto front is considerably flatter for large model lengths than the previous cases, namely that there is little improvement in accuracy when one makes the model more complex. We therefore choose an expression from the start of the flat region of the Pareto front, and find that the expression at length 46 is appropriate for our purposes.

As well as merging and removing superfluous parameters, we find that it is necessary to remove one term from this equation. The expression found by \operon{} contained a term which, when we evaluate the model for a $\Lambda$CDM universe, was proportional to $k$. Given that the expression from \citet{Bartlett_2024_linear} is a good approximation for $\Lambda$CDM universes, one would expect that $S(k,\bm{\theta})$ would be independent of $k$ for these cosmologies, and would simply scale the growth factor to correct for the mistakes obtained when using an approximate version, since the shape of $P_{\rm lin}(k, a, \bm{\theta})$ should be correct. Since the coefficient for this $k$-dependent term was small ($\mathcal{O}(10^{-4})$), we set this coefficient to zero to remove this term. 
After removing this term, we re-optimised the coefficients of the function using the training set. 
We find that the omission of this term has a negligible effect on the accuracy of the expression.

After performing these manipulations, we arrive at the expression
\begin{equation}
\label{eq:plin_S_result}
    \begin{split}
    10 & \log_{10} S \left( k, \bm{\theta} \right) \approx 
    - e_0 h - e_1 w_0 - \frac{e_2 m_{\rm \nu}}{\sqrt{e_3 + k^2}} - \frac{e_{4} h}{e_{5} h + m_{\rm \nu}} \\
    & + \frac{e_6 m_{\rm \nu}}{h \sqrt{e_7 + \left(e_8 \Omega_{\rm m}  + k \right)^2}} 
    + \frac{ e_{9} \Omega_{\rm b} - e_{10} w_0 - e_{11} w_a + \frac{e_{12} w_0 + e_{13}}{e_{14} w_a + w_0} }{\sqrt{e_{15} + \left( \Omega_{\rm m} + e_{16} \log{(-e_{17} w_0)} \right)^2}}.
    \end{split}
\end{equation}
when the coefficients $\bm{e}$ are tabulated in \cref{tab:plin_S_params}.

There are several interesting features in this approximation. Firstly, let us consider the $\Lambda$CDM limit, where we have $w_0 = -1$, $w_a=0$ and $m_{\nu} = 0$
\begin{equation}
    \label{eq:plin_S_lcdm}
    10 \log_{10} S_{\rm \Lambda CDM} = 
    - e_0 h + e_1 - \frac{e_4}{e_5} + 
    \frac{e_9 \Omega_{\rm b} + e_{10} + e_{12} - e_{13}}{\sqrt{e_{15} + \left( \Omega_{\rm m} + e_{16} \log e_{17} \right)^2}},
\end{equation}
which is not zero, and thus we have a correction even for these cosmologies. 
Given that $R(a, \bm{\theta})$ is unity at redshift zero so cannot correct for errors in $D_{\rm approx}$, this term gives a correction to the growth factor at $a=1$ due to using an approximation to this. This term provides a correction to the result of \citet{Bartlett_2024_linear}, as given in \cref{eq:lcdm_linear_F}. 
This $k$-independent correction adjusts the normalisation of the redshift-zero linear power spectrum, which arises since \citet{Bartlett_2024_linear} fitted the residuals between \camb{} and the \citeauthor{Eisenstein_1998} prediction, as returned by the \colossus{} package. In \colossus{}, the normalisation is set such that the result of computing \cref{eq:sigmaR} with the approximate power spectrum matches the desired $\sigma_8$. Given the limitations of the \citeauthor{Eisenstein_1998} approximation, this induces a multiplicative scaling to the power spectrum which can given errors of a few perfect on large scales (i.e. if one inferred $A_{\rm s}$ from the low-$k$ behaviour of the \colossus{} power spectrum, one would find a value which is wrong by a few percent). Although \cref{eq:lcdm_linear_F} removes this offset, since in this work we do not attempt to match $\sigma_8$ in this way, \cref{eq:plin_S_lcdm} acts to reintroduce this offset so that \cref{eq:lcdm_linear_F} yields the correct result.

Let us also consider the limiting behaviour of \cref{eq:plin_S_result}. On large scales ($k \to 0$), one observes that $S(k, \bm{\theta})$ tends to a constant, and thus introduces a multiplicative correction due to imperfections induced by the other approximations. It is essential that there is no $k$-dependence in this limit, since the power spectrum must be proportional to $k^{n_{\rm s} }$ on large scales, and this is already captured by \cref{eq:EH_nw}. We also note that on small scales ($k \to \infty$), $S(k, \bm{\theta})$ also tends to a (different) constant. This implies that the other terms in the approximation yield the correct $k$-dependence on small scales. We note that it is desirable that this term does not diverge even when extrapolated to large $k$, which is neither guaranteed nor analytically testable for a numerical emulator.

Finally, we note that, although we allowed such behaviour, \cref{eq:plin_S_result} does not contain any oscillatory terms. This is understandable since one would not expect the interaction between photons and baryons which give rise to the BAO features to depend on the nature of dark energy nor the sum of neutrino masses: these additional components should only affect the broad-band features of the power spectrum. This is indeed what we find, and further justifies our decision to find a correction to the $\Lambda$CDM result from \citet{Bartlett_2024_linear} rather than refit the whole equation, given the (hard to fit) oscillatory features are unchanged.

\begin{table}[]
    \caption{Best-fit parameters for the correction to the linear power spectrum given in \cref{eq:plin_S_result}. 
    }
    \centering
    \begin{tabular}{c|l|c|l|c|l}
    Param. & Value & Param. & Value & Param. & Value \\
    \hline\hline
$e_{0}$ & 0.2841 & $e_{6}$ & 0.0985 & $e_{12}$ & 0.1385 \\
$e_{1}$ & 0.1679 & $e_{7}$ & 0.0009 & $e_{13}$ & 0.2825 \\
$e_{2}$ & 0.0534 & $e_{8}$ & 0.1258 & $e_{14}$ & 0.8098 \\
$e_{3}$ & 0.0024 & $e_{9}$ & 0.2476 & $e_{15}$ & 0.019 \\
$e_{4}$ & 0.1183 & $e_{10}$ & 0.1841 & $e_{16}$ & 0.1376 \\
$e_{5}$ & 0.3971 & $e_{11}$ & 0.0316 & $e_{17}$ & 0.3733 \\
    \end{tabular}
    \label{tab:plin_S_params}
\end{table}

\subsubsection{Performance of combined approximation}

\begin{figure*}
    \centering
    \includegraphics[width=\textwidth]{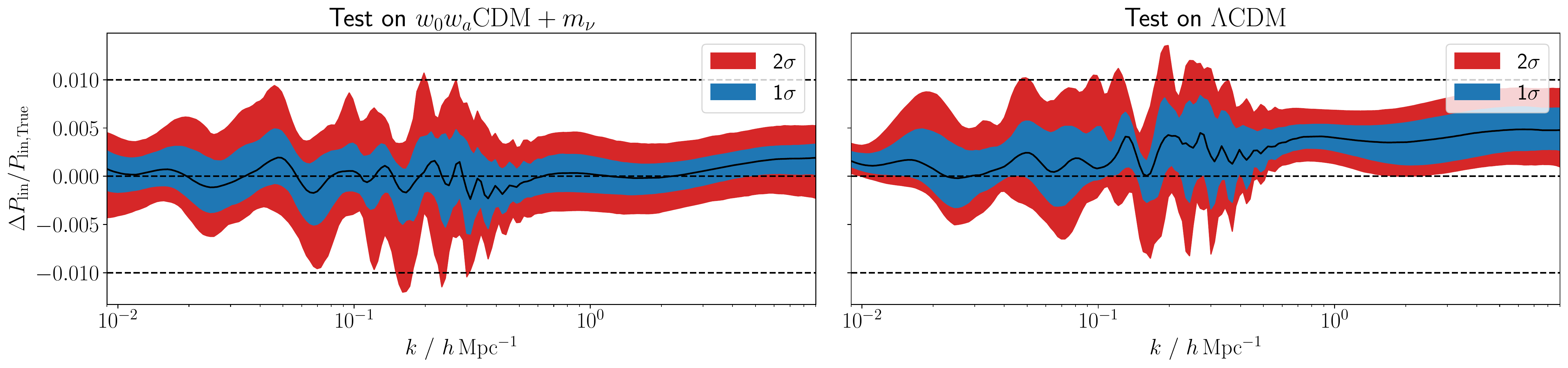}
    \caption{Distribution of fractional errors on $P_{\rm lin}(k, a, \bm{\theta})$ as a function of $k$ for the extended cosmological models considered in this work (left) and for $\Lambda$CDM (right). The bands given the 1 and $2\sigma$ values across the LH (\cref{tab:cosmo_par_prior}), with the dashed lines indicating 1\% error.
    When averaged over $k$, we find that our expressions have a root mean squared error of 0.28\% for the extended cosmology, and 0.41\% for $\Lambda$CDM.
    }
    \label{fig:linear_pk_predictions}
\end{figure*}

Now that we have verified that the individual components of the linear matter power spectrum are accurately approximated, we combine these approximations to assess the performance of $P_{\rm lin}(k, a, \bm{\theta})$ across the cosmologies and redshifts considered in this work.
Using a LH of 2000 points from the prior range in \cref{tab:cosmo_par_prior}, we compare the predictions from the expressions obtained in this work to the true linear power spectrum, as computed using \classcode.
We consider both the full parameter space and a separate LH where we restrict ourselves to $\Lambda$CDM only. The resulting distribution of errors as a function of $k$ are plotted in \cref{fig:linear_pk_predictions}, where the bands correspond to the 68\textsuperscript{th} and 95\textsuperscript{th} percentiles of errors across the test set at a given $k$.
Again, we consider 200 values of $k$ which are logarithmically spaced between $9 \times 10^{-3}$ and $9 \, h{\, \rm Mpc^{-1}}$.

We find that the root mean squared error at any given value of $k$ is much less than 1\%, and is 0.28\% for the full parameter space when averaged over $k$ and redshift. This increases to 0.41\% when we consider the $\Lambda$CDM sub-space, which is approximately a factor of 2 less accurate than the approximation of \citet{Bartlett_2024_linear}. We attribute this the fact that
zero neutrino mass is towards the edge of the range of our training set, where one would expect less accurate results. We explore this point further in \cref{sec:error distribution}. Nonetheless, we find that the $2\sigma$ error band when varied as a function of $k$ is almost always within 1\%. We assess the performance of this approximation further in \cref{sec:emulator_comparison}, where we compare to other emulators and fitting functions.

\section{Analytic emulator for the nonlinear power spectrum}
\label{sec:nonlinear_emulator}

Now that we have symbolic emulators for quantities related to the linear power spectrum, in this section we aim to produce an approximation for the nonlinear case.
To obtain such an expression, one must have access to $P_{\rm nl}(k, a, \bm{\theta})$ for a range of cosmological parameters and redshifts. 
As in \citet{Bartlett_2024_syren}, we obtained these by evaluating  \euclidemu{} \citep{Knabenhans_2021}, a method known for its speed and accuracy in predicting the ratio of the nonlinear to linear matter power spectrum. This approach was preferred over generating new $N$-body simulations, as it is significantly less computationally demanding. \euclidemu{} provides percent-level accuracy, allowing us to fit the emulated spectra with high precision and offering greater flexibility in selecting redshift values. Our results are validated against $N$-body simulations, as discussed in \cref{sec:emulator_comparison}.

Unlike \citet{Bartlett_2024_syren}, where corrections are applied to an existing \halofit{} formalism, our approach directly fits the nonlinear power spectrum, $P_{\rm nl}$, as a function of $k$, $a$, $\bm{\theta}$, and $P_{\rm lin}(k, a, \bm{\theta})$. By doing this, we avoid fitting additional \halofit{} variables, which can introduce further uncertainties. This method provides a more straightforward and potentially more accurate model for predicting the nonlinear power spectrum, since we do not have to correct for mistakes made by using a previous approximation. 

Similar to $S(k, \bm{\theta})$, we used a Latin hypercube sampling of 200 cosmologies with 200 logarithmically spaced $k$ values in the range $9 \times 10^{-3} - 9 \, h{\, \rm Mpc^{-1}}$ to generate the training set. For each cosmology, we generated the corresponding nonlinear power spectrum using \euclidemu{}. Since the nonlinear emulator requires $P_{\rm lin}$ as input, during training we used ground truth values computed by \classcode. However, for evaluating performance, we employed the $P_{\rm lin}$ emulator developed in \cref{sec:linear_emulator}.

\begin{figure}
    \centering
    \includegraphics[width=\columnwidth]{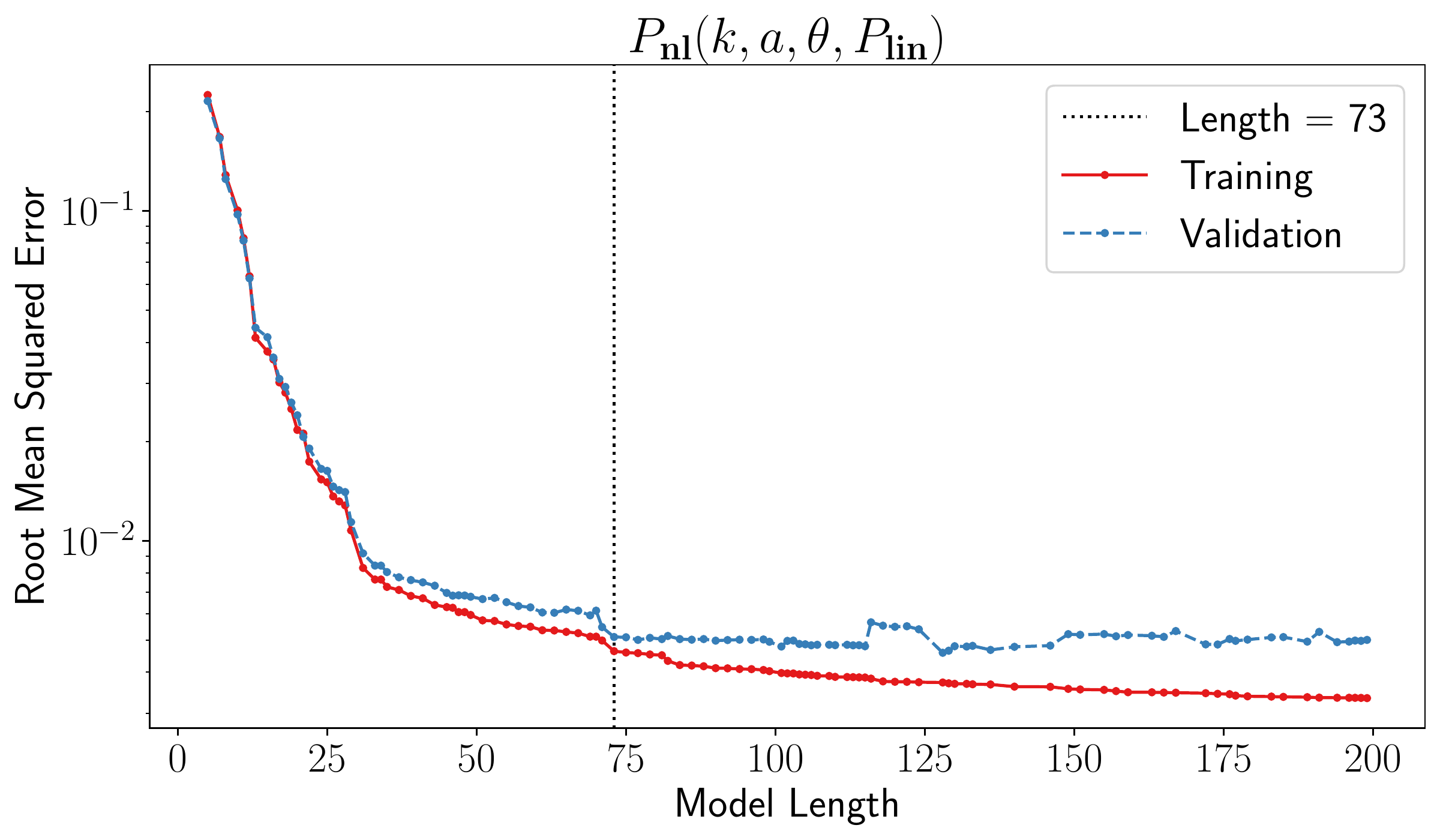}
    \caption{
    Same as \cref{fig:plin_pareto} but for $\log_{10}P_{\rm nl}(k, a, \bm{\theta},P_{\rm lin})$. 
    }
    \label{fig:pnl_pareto}
\end{figure}

We fit the nonlinear power spectrum $P_{\rm nl}(k, a, \bm{\theta},P_{\rm lin})$ using \operon. The model searches for expressions composed of the operators $+$, $-$, $\times$, $\div$, $\sqrt{\cdot}$, $\log$, and $\pow$. We found that excluding the analytic quotient operator improves the limiting behaviour, as it is easier to enforce the correct limits (discussed below) to the found expressions when this operator was removed. We set our target to $\log_{10}P_{\rm nl}$ and  transform the input $P_{\rm lin}$ to $\log_{10}P_{\rm lin}$ as well. We allow a maximum model length of 200, and we use the root mean squared error as the loss function, with $\epsilon = 10^{-3}$. After running for 24 hours on 56 cores, we obtain the Pareto front shown in \cref{fig:pnl_pareto}. From this, we select the expression with a model length of 73. The resulting equation is:

\afterpage{\FloatBarrier} 
\begin{widetext}
\begin{equation}
    \label{eq:pk_nl_fit}
    \begin{split}
    \log_{10}\tilde{P}_{\rm nl} = \log_{10}P_{\rm lin} + \frac{g_{0} k \left(g_{1} k\right)^{g_{2}\Omega_{m} - g_{3} \times 10^9 A_{\rm s} }}{\left(g_{4} k^{- g_{5}}- g_{6}\log_{10}P_{\rm lin}  \right)^{g_{7}\log_{10}P_{\rm lin}  + g_{8} w_{a}+ g_{9} w_{0} - g_{10} } + \left(g_{11} k^{g_{12}}+g_{13}\log_{10}P_{\rm lin}  - g_{14}\Omega_{m} \right)^{g_{15}a  - g_{16} n_{s}}} \\ + \frac{\left( g_{17}a - g_{18} \log_{10}P_{\rm lin}+ g_{19}\right)k }{g_{20}\Omega_{m}  + g_{21} k + g_{22} n_{s} - g_{23} + \left(g_{24}\log_{10}P_{\rm lin}  + g_{25} k^{g_{26}}\right)^{g_{27}a  - g_{28} n_{s}}}- g_{29} k -  \left(g_{30} k\right)^{\left(g_{31} k\right)^{- a g_{32}}},
    \end{split}
\end{equation}
\end{widetext}
\noindent
where the best-fit parameters are given in \cref{tab:pnl_params}. 
We denote this expression $\tilde{P}_{\rm nl}$ rather than ${P}_{\rm nl}$, since it is not our final approximation for the nonlinear power spectrum (see \cref{eq:pnl_fit_final} below).

\begin{table}[]
    \caption{Best-fit parameters for nonlinear power spectrum given in \cref{eq:pk_nl_fit}.}
    \centering
    \begin{tabular}{c|l|c|l|c|l}
    Param. & Value & Param. & Value & Param. & Value \\
    \hline\hline
$g_0$  & 0.2107 & $g_{11}$ & 0.9039 & $g_{22}$ & 1.4326 \\
$g_1$  & 0.0035 & $g_{12}$ & 0.0749 & $g_{23}$ & 1.8971 \\
$g_2$  & 0.0667 & $g_{13}$ & 0.0741 & $g_{24}$ & 0.0271 \\
$g_3$  & 0.0442 & $g_{14}$ & 0.1277 & $g_{25}$ & 0.9635 \\
$g_4$  & 1.2809 & $g_{15}$ & 27.6818 & $g_{26}$ & 0.0264 \\
$g_5$  & 0.2287 & $g_{16}$ & 24.8736 & $g_{27}$ & 22.9213 \\
$g_6$  & 0.1122 & $g_{17}$ & 0.6264 & $g_{28}$ & 71.1658 \\
$g_7$  & 4.3318 & $g_{18}$ & 0.3035 & $g_{29}$ & 0.0371 \\
$g_8$  & 1.1857 & $g_{19}$ & 0.6069 & $g_{30}$ & 0.0099 \\
$g_9$  & 3.3117 & $g_{20}$ & 0.7882 & $g_{31}$ & 210.3925 \\
$g_{10}$ & 14.2829 & $g_{21}$ & 0.4811 & $g_{32}$ & 0.2555 \\
    \end{tabular}
    \label{tab:pnl_params}
\end{table}

We find that this equation has several interesting properties. Firstly, one can analytically check the limiting behaviour as $k \to 0$. In this case, we find that $P_{\rm nl} \to P_{\rm lin}$ as $k \to 0$, consistent with our theoretical interpretation. It would not be possible to guarantee such behaviour using purely numerical emulators.
Second, we observe that $m_{\nu}$ does not appear explicitly; its dependence is only implicit through $P_{\rm lin}$.
This aligns with the results of \citet{Bayer_2022}, who found that all the information on the neutrino mass in the large scale structure is encoded in the linear power spectrum, and not the nonlinear evolution.

When evaluated on both the training and validation set, we observed that this equation by itself introduces a systematic offset compared to the \euclidemu{} results across all cosmologies and redshifts. 
This is a result of a sub-optimal search of the genetic algorithm, since it was unable to correct for this residual.
Therefore, we introduced a post-processing step to correct for this, where we fitted and subtracted the offset from the prediction. Again using \operon{} with the same settings as before, we fit for the offset as a function of $k$ only.
We find that the equation
\begin{equation}
    \label{eq:pnl_offset}
    {\rm offset}(k) = \frac{h_0 + \left(h_1 k - \cos\left(h_3 \cos\left(h_2 k\right)\right)\right) \cdot \cos\left(h_4 k\right) + \cos\left(h_5 k\right)}{-h_7 \cdot \log\left(h_6 k\right) + h_8 k},
\end{equation}
where $\bm{h}= \splitatcommas{[0.5787,   2.3485,  27.3829,  16.4236,  97.3766,  90.9764, 11.2046,  2447.2 , 11376.93]}$,
is sufficient to capture the remaining residual.
The final equation for nonlinear power spectrum is then given by
\begin{equation}
    \label{eq:pnl_fit_final}
    \log_{10}P_{\rm nl} =\log_{10}\tilde{P}_{\rm nl}-{\rm offset}(k).
\end{equation}
Note that this offset does not affect the limiting behaviour as $k \to 0$, since the offset tends to zero as $k \to 0$.
For large $k$, we find that this offset term does not exactly tend to zero, but oscillates around $h_1 / h_8 \approx 2 \times 10^{-4}$ with a period $\Delta k = 2 \pi/ h_4 \approx 0.06 \, h \, {\rm Mpc^{-1}}$. Such oscillations give a maximum correction to $\tilde{P}_{\rm nl}$ of 0.05\%, which is much smaller than the error on the power spectrum, and the variation with $k$ is much finer than the values of $k$ for which we reach this limit, which effectively averages out these oscillations. As such, we find that this offset becomes negligible as $k \to \infty$.

\begin{figure*}
    \centering
    \includegraphics[width=\textwidth]{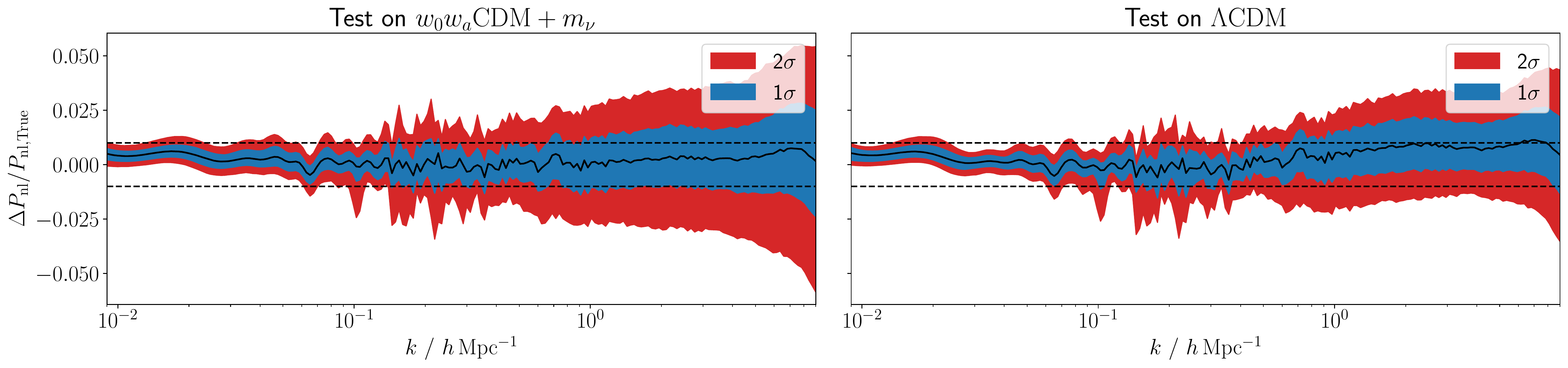}
    \caption{Same as \cref{fig:linear_pk_predictions} but for $P_{\rm nl}(k, a, \bm{\theta},P_{\rm lin})$, where we plot the errors relative to \euclidemu.
    When averaged over $k$, we find that our expressions have a root mean squared error of 1.3\% for the extended cosmology, and 1.1\% for $\Lambda$CDM.
    }
    \label{fig:nonlinear_pk_predictions}
\end{figure*}

Using a LH of 2000 points from the prior range in \cref{tab:cosmo_par_prior}, we compare our expressions' predictions to the outputs from \euclidemu{}, considering both the full parameter space and a restricted $\Lambda$CDM model. The results are given in \cref{fig:nonlinear_pk_predictions}. The root mean squared fractional error for the two cases is 1.3\% and 1.1\%, respectively.
We find that, in both cases, the error is dominated by the prediction at large $k$, where the $1\sigma$ difference between our result and \euclidemu{} can exceed 1\%, whereas the two emulators agree within 1\% before $k \sim 1 \, h \, {\rm Mpc^{-1}}$. Due to the inclusion of \cref{eq:pnl_offset}, when compared to Figure 5 of \citet{Bartlett_2024_syren}, we find that the residuals of our emulator at the BAOs are smaller and that the mean difference is much closer to zero.

\begin{figure}
    \centering
    \includegraphics[width=\columnwidth]{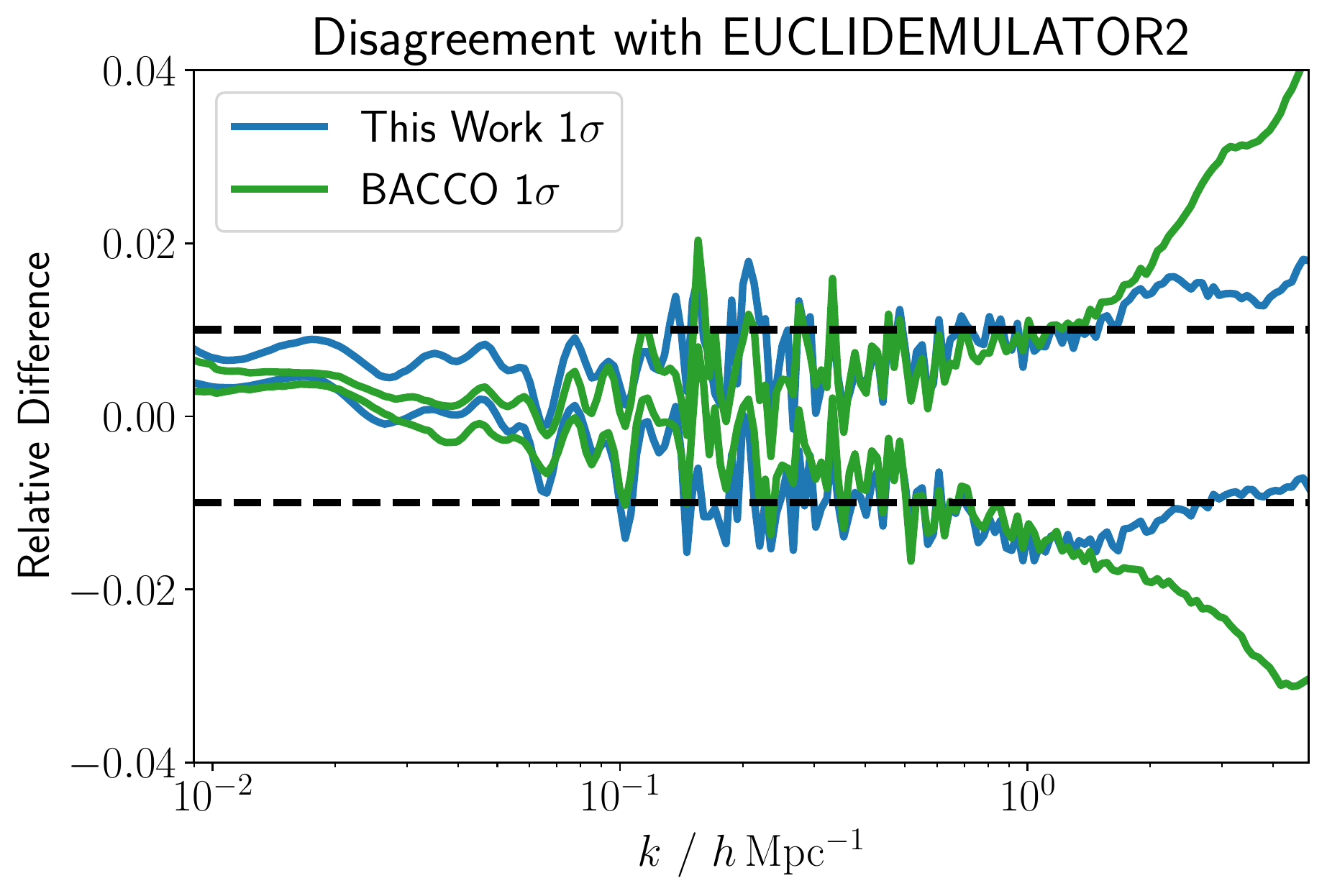}
    \caption{Distribution of fractional differences between different emulators. As a reference we use \euclidemu, and show the differences between this and the symbolic approximation obtained in this work and the \bacco{} emulator. We plot the $1\sigma$ differences across the LH used in this work, and the dashed lines correspond to 1\% agreement. The difference between our result and \euclidemu{} is comparable to the differences between the numerical emulators.}
    \label{fig:emulator_disagreement}
\end{figure}

To assess whether this level of disagreement is acceptable, we compare the difference between our symbolic emulator and \euclidemu{} to that between the numerical emulators \bacco{} and \euclidemu{} over the LH used in this work. In \cref{fig:emulator_disagreement} we plot the $1\sigma$ errors across the hypercube for both methods, where we see both our result and \bacco{} agree within 1\% as we go to $k\to 0$, as should be expected given than both methods produce sub-percent accurate linear power spectra. 
Around the scale of the BAOs, the level of disagreement with \euclidemu{} slightly increases to approximately 1\% for both emulators. As we increase $k$ further to $k > 1 \, h \, {\rm Mpc^{-1}}$, we find that our error is relatively stable at around 1\%, whereas the discrepancy between \euclidemu{} and \bacco{} rises dramatically to around 4\% by $k \sim 4 \, h \, {\rm Mpc^{-1}}$.
As such, we conclude that the level of disagreement between our approximation and \euclidemu{} is comparable to that between current numerical emulators, and thus we have reached a sufficiently accurate solution.
We investigate what this level of disagreement corresponds to for weak lensing measurements in \cref{sec:lensing}.

\section{Performance of emulators}
\label{sec:Performance}

\subsection{Comparison against other emulators}
\label{sec:emulator_comparison}

In this section we assess the performance of our emulator against those from the literature in terms of both run time and accuracy.
To assess accuracy, for $P_{\rm lin}(k, a, \bm{\theta})$, we compare the predicted power spectra to that computed by \classcode.
For the nonlinear case, however, it is less obvious what the standard against which we should compare is. In \cref{sec:nonlinear_emulator}, we calibrated our emulator against \euclidemu{} since this provided a computationally cheap way of generating a large number of spectra at a range of redshifts.
Now, however, we wish to compare against a $N$-body simulation, otherwise, by construction, \euclidemu{} would have zero error.
Therefore, as in \citet{Bartlett_2024_syren}, we benchmark the nonlinear emulators against the \quijote{} suite of simulations \citep{Quijote_sims}.

In particular, we test the accuracy against the following four sets simulations
\begin{enumerate}
    \item \textit{Fiducial:} This simulation has a $\Lambda$CDM cosmology with the \textit{Planck} 2018 cosmological parameters \citep{Planck_VI_2018}: $\Omega_{\rm m} = 0.3175$, $\Omega_{\rm b} = 0.049$, $h=0.6711$, $n_{\rm s} = 0.9624$, $\sigma_8=0.834$, $m_{\rm \nu} = 0$, $w_0 = -1$, $w_a = 0$.
    \item \textit{$M_{\rm \nu}^+$:} This simulation has a cosmological constant for dark energy, but contains three degenerate massive neutrino species of total mass $m_{\rm \nu} = 0.1 {\rm \, eV}$. All other parameters are the same as for the \textit{Fiducial} simulation.
    \item \textit{$w^+$:} In this simulation the dark energy equation of state is constant in time, but equal to -1.05 ($w_0 = -1.05$, $w_a = 0$). All other parameters are the same as for the \textit{Fiducial} simulation.
    \item \textit{$w^-$:} This is the same as the $w^+$ simulation, except $w_0 = -0.95$.
\end{enumerate}
All simulations contain $N^3=512^3$ particles within a cubic box of side length $L = 1 \, h^{-1} {\rm Gpc}$, and were run using the \textsc{gadget-iii} code \citep{Springel_2005}. The simulations containing neutrinos contain an additional $512^3$ neutrino particles.
We make use of the publicly available power spectra for these simulations for $k \in {[0.02, 0.5]} \, h \, {\rm \, Mpc}^{-1}$, since we expect these to be converged in this range \citep{Bartlett_2024_syren}.
To remove the effects of cosmic variance (i.e. to obtain an estimate of the ensemble mean power spectra), we average these spectra across all available simulations in each suite (15,000 for the \textit{Fiducial} simulation and 5,000 for the others).

\begin{figure*}
    \centering
    \includegraphics[width=\linewidth]{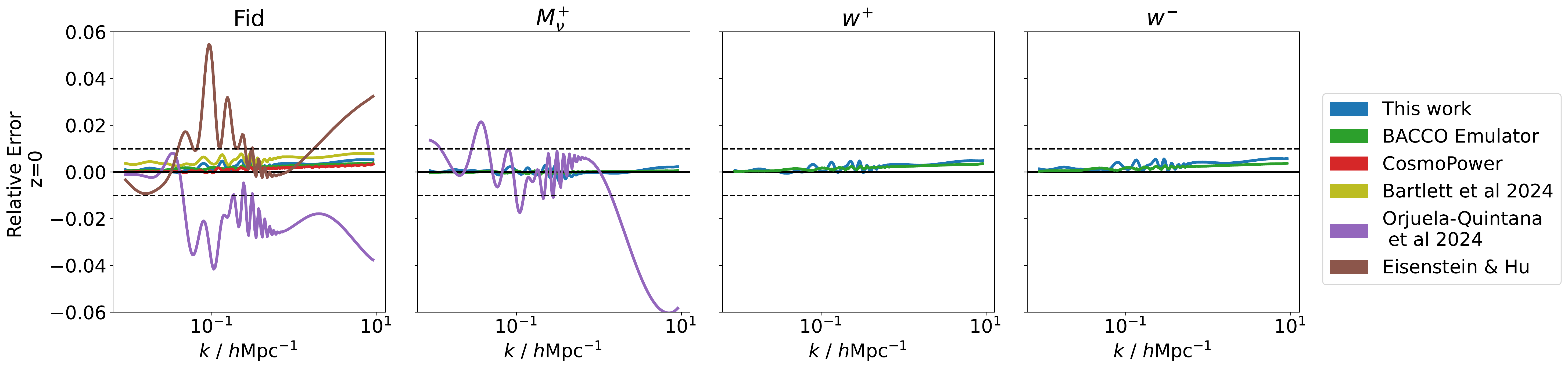}
    \caption{Fractional error on the linear power spectrum as a function of wavenumber, $k$, for various emulators at redshift 0. We apply each emulator only to test data that falls within its defined parameter space. 
    } 
    \label{fig:plin_emulators_comparison}
\end{figure*}

We first compare various linear power spectrum emulators across the four cosmologies. The emulators considered are
\bacco{} \citep{Angulo_2021,Arico_2021,Zennaro_2023}, 
\cosmopower{} \citep{SpurioMancini_2022}, 
and three equation-based models from previous works: 
\cite{Eisenstein_1998}, \cite{Bartlett_2024_linear}, and \cite{Orjuela-Quintana_2024}. 
Each emulator is applied only to cosmologies that fall within its defined parameter space. Since \cosmopower{} can only be evaluated at fixed 
\(k\) values, we apply linear interpolation to enable comparisons on arbitrary \(k\). \cref{fig:plin_emulators_comparison} shows the relative error as a function of wavenumber \(k\) across different cosmologies at redshift 0. 

Among the emulators considered, only \bacco{} and this work can be applied to all four cosmologies, and they show similar errors across the board, which is much less than 1\% for all $k$.
For the fiducial cosmology, the numerical emulators (\bacco{} and \cosmopower), the symbolic approximation of \citet{Bartlett_2024_linear}, and this work have power spectra which remain within 1\% for all $k$, suggesting these all have the requisite accuracy for modern cosmological analyses.
This is not the case for the \citet{Eisenstein_1998} and \citet{Orjuela-Quintana_2024} formulae, whose errors can reach several percent, and are particularly large near the BAOs. Although they have the correct limiting behaviour by construction as $k \to 0$, their errors are much larger than what is required for current and future surveys \citep{Taylor_2018}.

\begin{figure*}
    \centering
    \includegraphics[width=\linewidth]{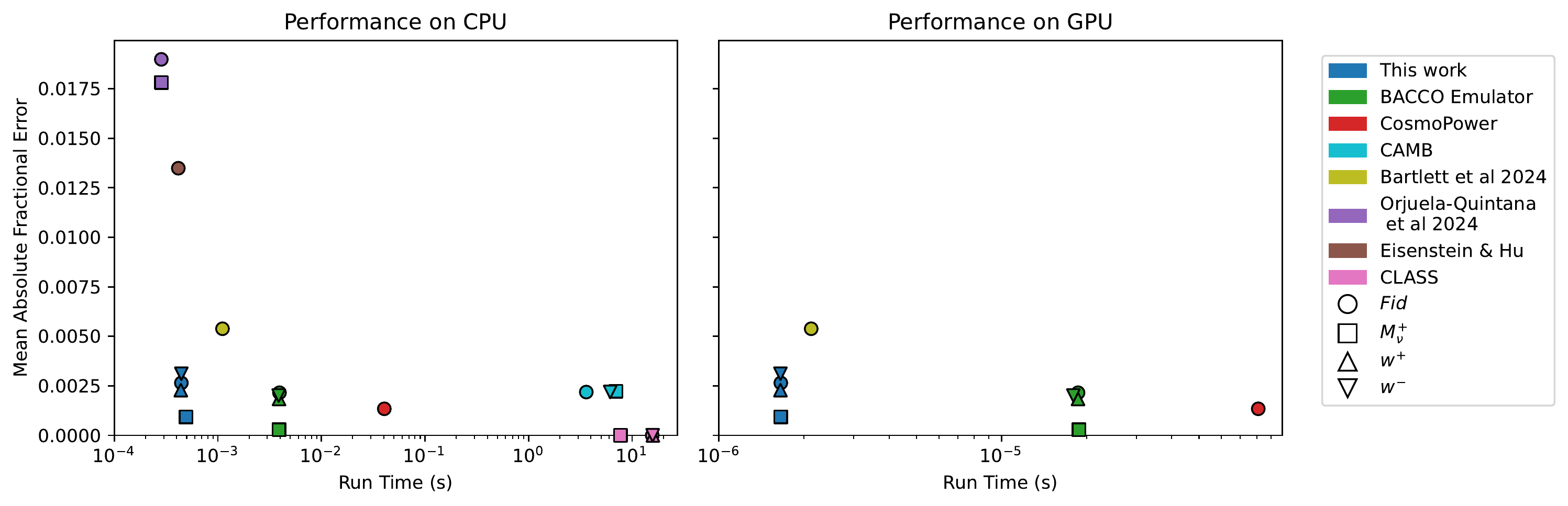}
    \caption{Mean absolute fractional error (averaged over 200 logarithmically spaced $k \in [ 9 \times 10^{-3}, 9 ] \, h \, \text{Mpc}^{-1}$)  with respect to \classcode{} against run time for various linear power spectrum emulators at four cosmologies and redshift 0. Some emulators are tested on both CPU (left panel) and GPU (right panel). The different colours refer to the emulator used, whereas each cosmology (defined in \cref{sec:emulator_comparison}) is denoted by a different symbol.
    } 
    \label{fig:pln_run_time_comparison}
\end{figure*}

To assess the relative runtime of the emulators, we evaluated the linear power spectrum at redshift 1 for the four cosmologies mentioned above, performing \(10^3\) evaluations on an Intel Xeon E5-4650 CPU. We report the mean execution time per evaluation. The calculation used 200 logarithmically spaced values of \(k\) in the range \(k = 9 \times 10^{-3} - 9 \, h \, \text{Mpc}^{-1}\). We assumed the availability of either \(A_{\rm s}\) or \(\sigma_8\), depending on the emulator's requirements.

The results of this timing test are shown in \cref{fig:pln_run_time_comparison}, where we plot accuracy -- computed as the mean absolute fractional error (averaged over \(k\)) at redshift 1 across the four cosmologies -- against runtime. Our emulator exhibits similar errors (within approximately 0.3\%) to \bacco{} and \cosmopower, while being 8 and 80 times faster than them, respectively, on the CPU. Additionally, we include the runtime of two simulators, \classcode{} and \camb{}, showing their differences in simulating the linear power spectrum. We find that our emulator obtains similar errors as there are discrepancies between the two codes (note that by definition \classcode{} has zero error in \cref{fig:pln_run_time_comparison}), but is a factor $3\times10^4$ faster.
Although they are of comparable speed to \syren{} and this work, the alternative analytic approximations of \citet{Eisenstein_1998,Eisenstein_1999,Orjuela-Quintana_2024} are significantly less accurate, with mean errors much greater than a percent, and maximum errors of several percent (see \cref{fig:plin_emulators_comparison}).
We find that the result of this work is faster than \citet{Bartlett_2024_linear}, which is to be expected given that we have removed the integral associated with using the \colossus{} package (see \cref{sec:plin_S}).

Two of the emulators (\bacco{} and \cosmopower{}) are implemented in TensorFlow, making them more efficient for GPU execution. To account for this, we compare a PyTorch implementation of our emulator and of \syren{} with \bacco{} and \cosmopower{}.  We run these four emulators on an NVIDIA Tesla V100 GPU, using a batch size of 2000 for each emulator, and compare their performance. The results are shown in the right panel of \cref{fig:pln_run_time_comparison}, where we see that all emulators are significantly faster. However, due to its simplicity, we still find that our emulator is an order of magnitude faster than both \cosmopower{} and \bacco{}, with an evaluation time of less than $2\,{\rm \mu s}$. Again, we find \syrenplus{} to be faster than \syren{}.

We also conducted similar comparisons with several nonlinear emulators:
\euclidemu{} \citep{Knabenhans_2021}, 
\bacco{} \citep{Angulo_2021,Arico_2021,Zennaro_2023}, 
\cosmopower{} \citep{SpurioMancini_2022}, 
\syren{} \citep{Bartlett_2024_syren}, 
and the \halofit{} \citep{Smith_2003,Bird_2012,Takahashi_2012} implementation in \classcode{}.
\euclidemu{} and \bacco{} are trained to predict the ratio of the power spectrum of $N$-body simulations to the linear one, whereas \cosmopower{} replicates the results of \hmcode{} \citep{Mead_2021}, which is a semi-analytic approach based on the halo model.
These emulators were compared against \quijote{} simulations across four suites and at three different redshifts (0, 0.5 and 1.5). 
Since \cosmopower{} has two additional free parameters -- $c_{\rm min}$, which describes the minimum halo concentration, and $\eta_0$, which determines the halo bloating -- we optimise these parameters to minimise the mean squared error between the prediction of $\log_{10} P_{\rm nl}$ and that from \quijote. The timing results for \cosmopower{} do not include this optimisation time and refers to a single pass once $c_{\rm min}$ and $\eta_0$ are known.

\begin{figure*}
    \centering
    \includegraphics[width=\linewidth]{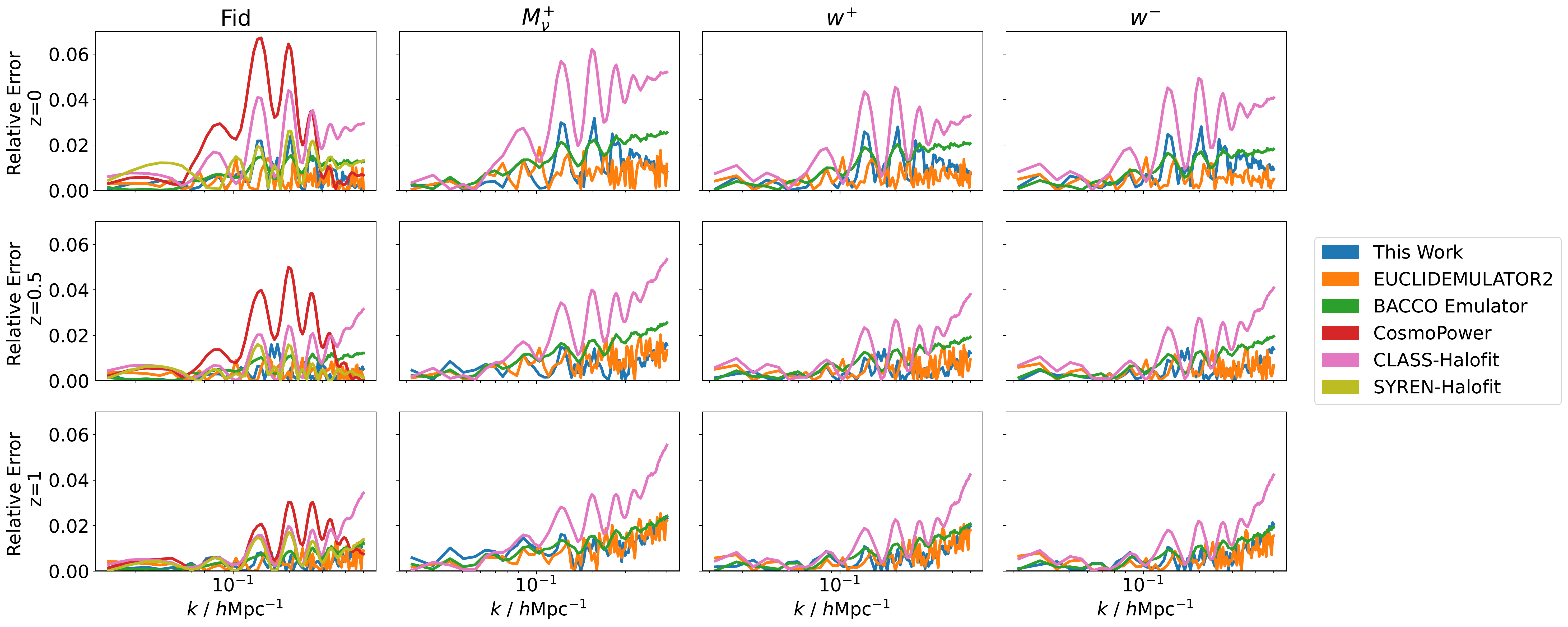}
    \caption{Fractional error on the nonlinear power spectrum as a function of wavenumber, $k$, for various emulators at redshifts 0, 0.5 and 1. We apply each emulator only to test data that falls within its defined parameter space. 
    } 
    \label{fig:pnl_emulators_comparison}
\end{figure*}

\begin{figure*}
    \centering
    \includegraphics[width=\linewidth]{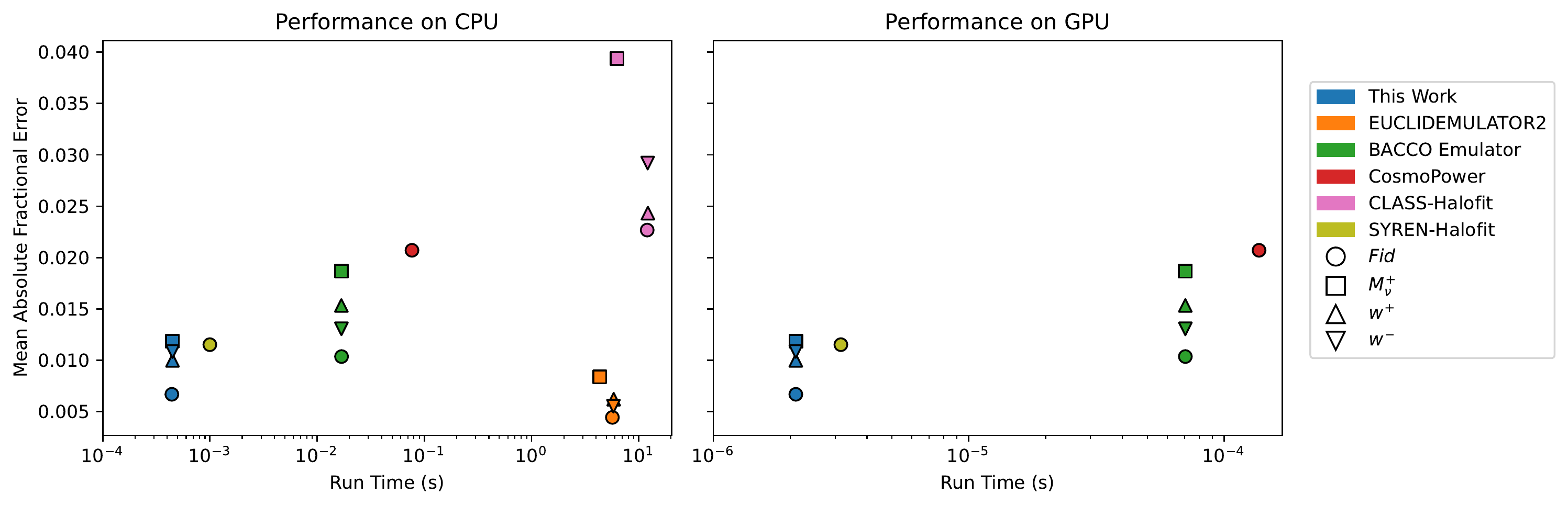}
    \caption{Same as \cref{fig:pln_run_time_comparison}, but now comparing nonlinear emulators with the \quijote{} simulations and averaging over $k \in {[0.02, 0.5]} \, h \, {\rm \, Mpc}^{-1}$.
    } 
    \label{fig:pnl_run_time_comparison}
\end{figure*}

The results, shown in \cref{fig:pnl_emulators_comparison} as absolute fractional errors, indicate that our emulator performs similarly to \euclidemu{} and \bacco{} across all suites. \syren{} is applicable only to the fiducial case and shows comparable errors to our method. In contrast, \cosmopower{} and \classcode-\halofit{} generally exhibit larger errors compared to our results.
\cref{fig:pnl_run_time_comparison} displays the timing results against the mean absolute fractional error. Our emulator is shown to be 40 times faster than \bacco{} and $10^4$ times faster than \euclidemu{} on a CPU, all while maintaining comparable error levels. As before, we also compare a PyTorch implementation of our emulator with the two TensorFlow-based emulators in the right panel. We find that our emulator is 65 times faster than \cosmopower{} and 33 times faster than \bacco{}, and again takes approximately $2\,{\rm \mu s}$ to evaluate.

\subsection{Error distribution}
\label{sec:error distribution}

Our results demonstrate that our linear emulator can achieve a root mean squared fractional error of less than 0.3\%, and this is around 1\% for the nonlinear emulator on the test set, which is sampled from our wide prior range. However, in applications such as inference problems, accuracy around the true cosmology is often more critical. To illustrate how the error varies with the distance from our current best-fit cosmology, we use cosmological constraints from the \textit{Planck} +  BAO + RSD + SN + DES chains of \citet{eboss20}.
Specifically, we fit the marginal posteriors using a multivariate Gaussian distribution and define the distance of a given model from the fiducial one as the probability density provided by the Gaussian. For interpretability, we express this distance in terms of $\sigma$ values. We generate several new test sets within 1, 3, 5, 7, and 9$\sigma$, sampling parameters to be uniformly distributed in this range. Additionally, we compare these results to errors across the entire prior space.

The \cite{eboss20} study does not provide posteriors that include all three of the $w_{0}$, $w_{a}$ and $m_{\nu}$ parameters. Consequently, we cannot directly define a distance metric in our parameter space using their results. Therefore, we consider two subspaces of the 8-dimensional prior space: the $\Lambda$CDM parameters ($A_{s}$, $\Omega_m$, $\Omega_b$, $h$, $n_s$) combined with $w_{0}$ and $w_{a}$, and the  $\Lambda$CDM parameters combined with $m_{\nu}$ and $w_{0}$. These two posteriors define the distance in their respective subspaces, while we fix the remaining parameters as assumed by \cite{eboss20}.

\begin{figure}
    \centering
    \includegraphics[width=\columnwidth]{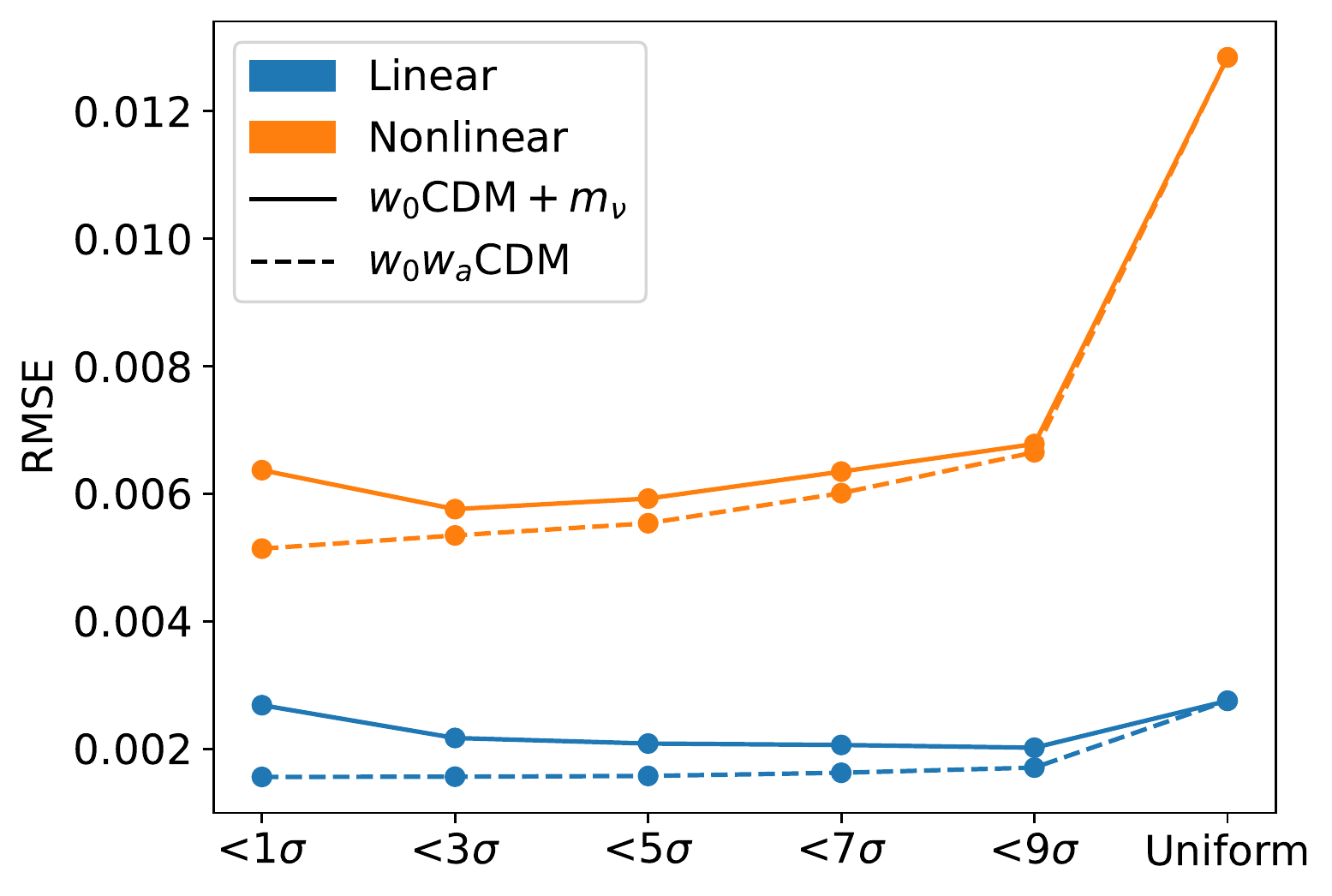}
    \caption{ Distribution of errors of the linear (blue) and nonlinear (orange) emulator as a function of distance from the best-fit model in the $w_{0}w_{a}$CDM subspace (dashed lines) and the $w_{0}$CDM+$m_{\nu}$ subspace (solid lines). The $x$-axis represents the maximum $\sigma$ value used for generating the test set. The "Uniform" case corresponds to generating test samples by uniformly sample parameters in the whole prior range. 
    }
    \label{fig:error_dist}
\end{figure}

\cref{fig:error_dist} displays the distribution of errors as a function of distance from the best-fit model. 
For the $w_{0}w_a$CDM model, we find that the error decreases as we approach the centre of the posterior, such that the error on the linear power spectrum is below 0.2\% within $1\sigma$ of the posterior mean, and this is below 0.5\% for the nonlinear case, which is over a factor of two better than across the full prior range. 

In the $w_{0}$CDM+$m_{\nu}$ subspace, although the error is smaller near the centre compared to the uniformly sampled test set, it increases slightly when approaching the $m_{\nu}=0$ point. This increase is due to the boundary effect in training: the lower limit for $m_{\nu}$ in the training set is $m_\nu = 0$, which is problematic since we find empirically that our predictions typically become less accurate near the training set boundaries. For example, we see in \cref{fig:linear_pk_predictions} that the massless neutrino model has slight bias when we assume a cosmological constant for dark energy.
Nonetheless, in the $w_{0}$CDM+$m_{\nu}$ subspace, the error within $9\sigma$ of the posterior mean is below  0.3\% for the linear power spectrum and still a factor of 2 smaller than for the full prior range when considering the nonlinear emulator.

\subsection{Performance for weak lensing analysis}
\label{sec:lensing}

In \cref{sec:nonlinear_emulator} and \cref{sec:emulator_comparison}, we found that the discrepancy between other emulators or $N$-body simulations and our approximations is approximately at the percent-level for the nonlinear matter power spectrum, and that this difference is greatest at large $k$.
In this section, we wish to see what this 1\% disagreement translates into when we move from just $P(k)$ to precisely analysing cosmological observations from an upcoming large scale structure survey and to determine whether this is acceptable or not.

We focus on the upcoming weak lensing observations from the Rubin observatory's Legacy Survey of Space and Time (LSST) \citep{LSST_2009}, which directly probes the matter power spectrum. We assess the performance of the model for two stages of the planned observations, year 1 (denoted as LSST Y1) and year 6 (denoted as LSST Y6). Following \cite{DESC:SRD:2018}, we assume that LSST Y1 will cover a sky area of $f_{\rm sky} = 0.3$ and that this will increase to $f_{\rm sky} = 0.4$ for Y6 analyses. 

For these observations, the source samples are expected to follow a distribution given by
\begin{equation}
    n^{\rm tot}_{\kappa}(z) \propto z^2 \exp[-(z/z_{\rm pz; 0})^{\alpha_{\rm pz}}],
\end{equation}
which is normalised by the effective number density $\bar{n}^{\rm tot}_{\kappa}$. For LSST Y1, we assume $\bar{n}^{\rm tot}_{\kappa} = 11.2$, $\alpha_{\rm pz} = 0.87$, and $z_{\rm pz; 0} = 0.191$; for LSST Y6, we assume $\bar{n}^{\rm tot}_{\kappa} = 23.2$, $\alpha_{\rm pz} = 0.798$, and $z_{\rm pz; 0} = 0.178$ \citep{DESC:SRD:2018}. 
This total distribution is divided into five tomographic bins with equal number densities for each bin $i$, $\bar{n}^{i}_{\kappa} = \bar{n}^{\rm tot}_{\kappa}/5$. For shape noise, we expect $\sigma_e = 0.26/{\rm component}$.

Given these source distributions, we compute the tomographic cosmic shear power spectra as a function of angular scale $\ell$ as
\begin{equation}
    \label{eq:Cl2h}
    C^{ij}_{\ell} = \int_{z_{\rm{min}}}^{z_{\rm{max}}} dz \frac{dV}{dz d\Omega} \frac{W^{i}_{\kappa}(z)}{\chi^2} \ \frac{W^{j}_{\kappa}(z)}{\chi^2} \ P(k,z),
\end{equation}
where, $i,j$ are the indices of tomographic bins, $k = (\ell + 1/2) / \chi$, $\chi$ is the comoving distance to redshift $z$, the integral is performed between $z_{\rm min}=0$ and $z_{\rm max} = 3.0$, and $W^{i}_{\kappa}(z)$ is the lensing efficiency given by:
\begin{equation}
    W^{i}_{\kappa}(z) = \frac{3 H_0^2 \Omega_{\rm m}}{2 c^2} \frac{\chi}{a(\chi)} \int_{\chi}^{\infty} d\chi' n^{i}_{\kappa}(z(\chi')) \frac{dz}{d\chi'}\frac{\chi' - \chi}{\chi'}.
\end{equation}
Here, $n^{i}_{\kappa}$ represents the normalised redshift distribution of the source galaxies corresponding to the tomographic bin $i$.

Next, we calculate the covariance for the auto- and cross-correlations of cosmic shear power spectra for the five tomographic bins. In this work, we assume a Gaussian covariance, as described in \cite{Takada:2004:MNRAS:}. Note that we ignore the contributions non-Gaussian contributions arising from the tri-spectrum of matter fluctuations \citep{Sato_2013} and super-sample covariance \citep{Krause:2017:MNRAS:}. The super-sample covariance has been shown to be non-negligible for current and future-generation weak lensing surveys \citep{Barreira_2018}. However, we remain conservative in our analysis choices and ignore their contributions here when quantifying the accuracy of our emulator. 

To compare the effects of using different matter power spectra, for a LSST-like survey we need to be able to evaluate $P(k, z)$ for approximately $k \lesssim 10  h^{-1} \, {\rm Mpc^{-1}}$ and for $z < 3$.
Given the coarse spacing of the snapshots of the \quijote{} simulation, unlike in \cref{sec:emulator_comparison}, we choose not to use this as a benchmark as it is not clear that the interpolation required between redshifts would be accurate. Instead, we compare the predictions of different emulators, to see if their differences result in significantly different $C_\ell$'s.
Given the large discrepancies between \classcode-\halofit{} and \cosmopower{} when compared against \quijote{} (\cref{fig:pnl_emulators_comparison}), we choose not to use these codes in our comparison, but consider those with approximately percent-level agreement: \syren, \syrenplus, \bacco{} and \euclidemu. In fact, we cannot use \bacco{} since this is only trained up to $z=1.5$, and thus cannot be used for higher-redshift surveys like LSST.
For the three remaining codes, we assume the flat $\Lambda$CDM cosmology $h = 0.6766$, $\Omega_{\rm m} = 0.3111$, $\Omega_{\rm b} = 0.0490$, $\sigma_8 = 0.8102$, $n_{\rm s} = 0.9665$ \citep{Planck_VI_2018} with massless neutrinos.
To compute $C_{\ell}$, we evaluate $P(k,z)$ on a $128\times128$ two-dimensional grid, uniform in $\log_{10} k$ and $z$, where $5 \times 10^{-3} \leq k \ / \ h^{-1} \, {\rm Mpc^{-1}} \leq 9$ and $0 < z < 3$, which we use to numerically perform the integral \cref{eq:Cl2h}.

\begin{figure*}
    \centering
    \includegraphics[width=\textwidth]{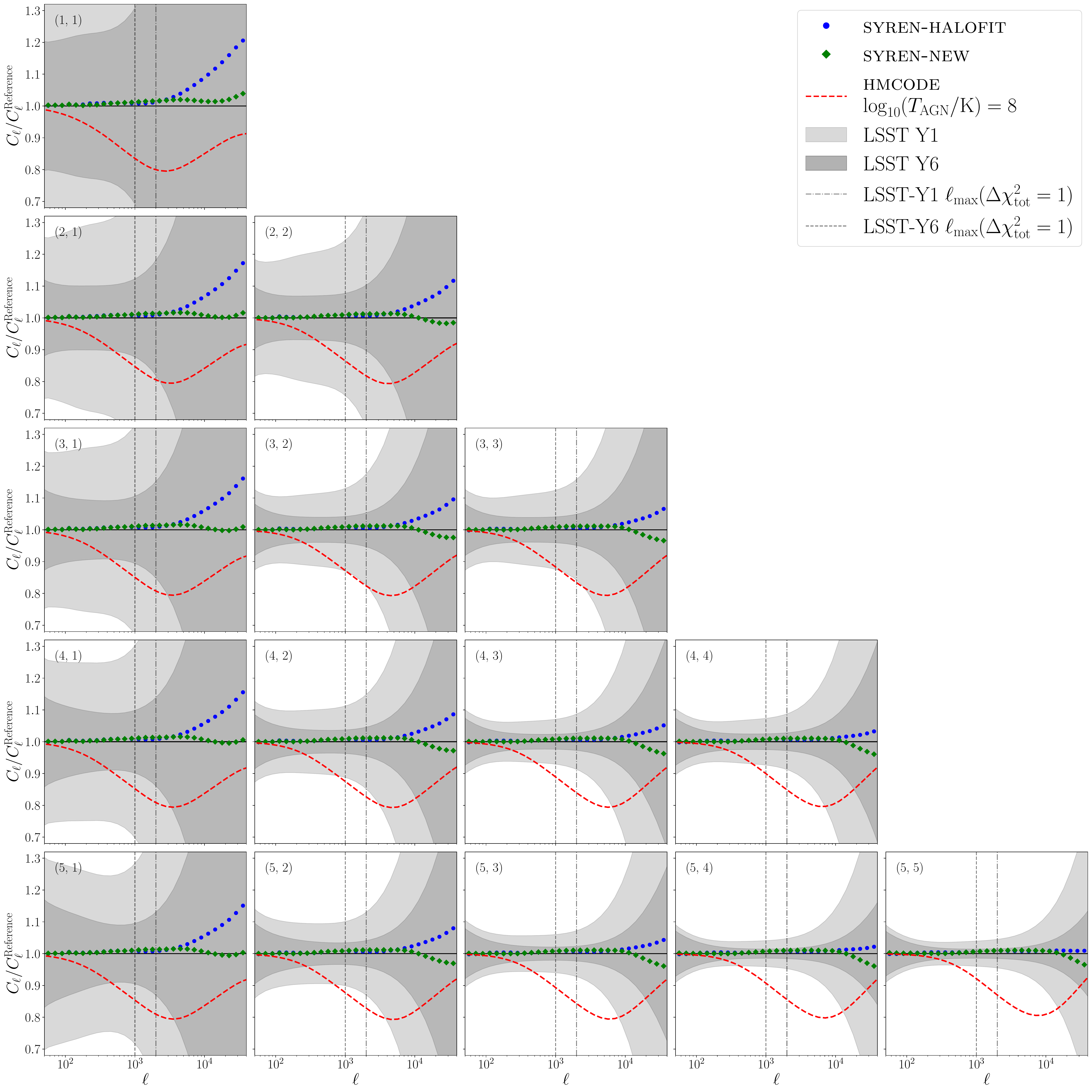}
    \caption{Ratio of shear power spectra, $C_\ell$ (\cref{eq:Cl2h}), computed with \syren{} and \syrenplus{} compared to that computed with \euclidemu{} for LSST. The light grey bands indicate the expected uncertainty for year 1 (Y1), whereas the dark bands are for year 6 (Y6). The numbers in the top left of each panel indicate the redshift bins between which we compute the $C_\ell$'s, such that the diagonal panels are auto-spectra and the off-diagonal ones are cross-spectra. 
    To indicate the importance of baryonic effects, in red we plot the ratio of the \hmcode{} prediction in the presence of baryons (with $\log_{10} (T_{\rm AGN} \ / \ {\rm K}) = 8$) to the dark-matter-only \hmcode{} prediction.
    With vertical lines, we show the scales for which the $\Delta \chi^2_{\rm tot}=1$ between our emulators and \euclidemu{} for full datavectorfor for both LSST Y1 and Y6.
    The discrepancy between the emulators is small compared to the LSST errors and especially compared to the impact of baryonic effects for the angular scales shown.
    }
    \label{fig:lensing}
\end{figure*}

The resulting $C_\ell$'s are plotted in \cref{fig:lensing}, where we plot the ratio between the predictions from different emulators and that of \euclidemu{}, for $\ell < 4 \times 10^4$. For reference, we plot the expected measured errors for Y1 and Y6 in bands. Qualitatively, we see that the differences between the predictions are always within the $1\sigma$ band for both Y1 and Y6 for these angular scales, indicating that the percent-level disagreement for $P(k)$ does not lead to significant differences on $C_\ell$.
To estimate the range of angular scales for which the emulators do not yield significantly different results, we compute the total $\Delta \chi^2$ between \euclidemu{} and the symbolic emulators when including all scales up to some maximum $\ell < \ell_{\rm max}$.
To ensure that this $\Delta \chi^2$ remains less than unity, for \syrenplus{} one would have to use $\ell_{\rm max} = 2000$ for Y1 data and $\ell_{\rm max} = 1000$ for Y6, as indicated by the vertical lines in \cref{fig:lensing}.
We note that a full cosmological inference is required to quantify the impact of these $\Delta \chi^2$ on the parameter biases which will depend on specific analysis choices and observational systematics, and is outside the scope of this study. \footnote{However, as the residuals between our emulators and \euclidemu{} have a complicated shape, we expect a negligible impact on the $S_8$ parameter that weak lensing mainly constrains \citep{Jain:1997:ApJ:} and is sensitive to the scale-independent amplitude of the power spectrum.} 

On small scales, the effects of baryonic effects (e.g. astrophysical feedback from AGNs) and intrinsic alignments on the lensing power spectra are expected to be significant and which are poorly understood (e.g. see \cite{Krause:2021:arXiv:} and references there-in). \footnote{Note that the Dark Energy Survey uses a range of scale-cuts corresponding to approximately $\ell \lesssim 500$ \citep{Doux_2022} to mitigate these effects, which is much stricter than our $\ell_{\rm max}$.} To estimate their importance, we use the halo model-based formalism, \hmcode{} \citep{Mead_2021}, developed to capture the impact of baryonic feedback using an effective $T_{\rm AGN}$ parameter which controls the energy output from AGNs. In \cref{fig:lensing} we plot the ratio of $C_\ell$ between the \hmcode{} predictions, when either excluding or including baryonic effects, where we choose an AGN temperature $T_{\rm AGN} = 10^8 \, {\rm K}$ as expected from various hydrodynamical simulations \citep{LeBrun:2017:MNRAS:, McCarthy:2018:MNRAS:}.
We see that this ratio is much larger than the differences between the dark-matter-only emulators, and thus conclude that the dominant source of uncertainty when used in a weak lensing analysis is not due to the choice of emulator, but due to baryonic effects. 
As such, we advocate for the use of \syrenplus{} in such an analysis, given its speed and accuracy compared to other emulators (\cref{fig:pnl_run_time_comparison}).

\section{Conclusion}
\label{sec:Conclusion}

With current and future surveys poised to probe the massess of neutrinos and the nature of dark energy, rapid and accurate predictions of their effects on the linear and nonlinear evolution of matter in the Universe is essential. In this paper we have used symbolic regression to obtain analytic approximations to the linear and nonlinear matter power spectra as a function of redshift and cosmology in the presence of massive neutrinos and dynamical dark energy.

We obtained a succinct correction to previous symbolic approximations to the linear matter power spectrum which is accurate to within 0.3\% across a wide range of cosmologies yet is $3\times10^4$ times faster than using a Boltzmann code when both are run on a CPU. A further factor of 200 speed-up is possible when running on a GPU. Similarly, we produced a formula to compute the derived parameter $\sigma_8$ as a function of $A_{\rm s}$ and other cosmological parameters, which does not require one to obtain or integrate the linear power spectrum, and is accurate to 0.1\%.
Analogous to \halofit, we found a fitting formula to predict the nonlinear matter power spectrum given the linear one, redshift and cosmological parameters.
This approximation matches the predictions of \euclidemu{} \citep{Knabenhans_2021} to within 1.3\% across a broad range of parameter space, redshift and wavenumbers, which is comparable to the level of discrepancy between different numerical emulators. The error is halved when considering cosmologies within the $9\sigma$ region of current cosmological constraints \citep{eboss20}, so we have sub-percent agreement for cosmologies of physical interest.
We find that the difference does not lead to significantly different predictions for modern, complex, multi-redshift bin, tomographic weak lensing surveys.
We validated these predictions against $N$-body simulations, finding similar accuracy to numerical emulators while being over an order of magnitude faster, both on a CPU and GPU.

Unlike approaches such as \hmcode, our \halofit-like procedure does not include the effects of baryonic feedback, which become important on the smallest scales considered in this work. Future work should therefore be dedicated to obtaining baryonic corrections to our approximations so that one can marginalise over or infer these effects in cosmological analyses.
We also note that here, as in \citet{Bartlett_2024_syren}, we chose to fit our models to the predictions of numerical emulators, which were themselves trained on $N$-body simulations since this provided an inexpensive method for producing power spectra. There is no fundamental reason to do this, and indeed one could limit the performance of the symbolic expression if the numerical emulator were not sufficiently accurate. 
As such, in future work it may be desirable to directly fit the symbolic models to simulations rather than use an intermediary numerical emulator.

Symbolic methods continue to demonstrate an ability to produce fitting formulae that are significantly faster to evaluate than numerical approaches without being less accurate.
Such formulae are interpretable, and the correct limiting behaviour may readily be enforced.
These advantages, coupled with the lack of reliance on particular programming languages or packages, 
will make symbolic methods valuable for a range of emulation tasks in astrophysics and cosmology.

\section*{Acknowledgements}

We thank 
Marco Bonici, 
Lukas Kammerer
and
Gabriel Kronberger
for useful comments and suggestions. 
CS is supported by the National SKA Program of China (grant No. 2020SKA0110401) and NSFC (grant No. 11821303).
DJB and SP are supported by the Simons Collaboration on ``Learning the Universe.''
BDW acknowleges the DIM-ORIGINES-2023, \textit{Infinity Next} grant.
HD is supported by a Royal Society University Research Fellowship (grant no. 211046).
PGF acknowledges STFC and the Beecroft Trust.
This work has made use of the Infinity Cluster hosted by Institut d'Astrophysique de Paris; we thank Stephane Rouberol for running this cluster smoothly for us. Some of the computations reported in this paper were performed using resources made available by the Flatiron Institute. The Flatiron Institute is supported by the Simons Foundation. 

For the purposes of open access, the authors have applied a Creative Commons Attribution (CC BY) licence to any Author Accepted Manuscript version arising.

The data underlying this article will be shared on reasonable request to the corresponding authors.
We provide implementations of our emulators at \url{https://github.com/DeaglanBartlett/symbolic_pofk}.

\bibliographystyle{aa} 
\bibliography{references} 

\begin{appendix} 

\section{Most accurate analytic expression found for $\sigma_{8}$}
\label{sec:most_accrurate_sigma8}

\begin{figure}
    \centering
    \includegraphics[width=\columnwidth]{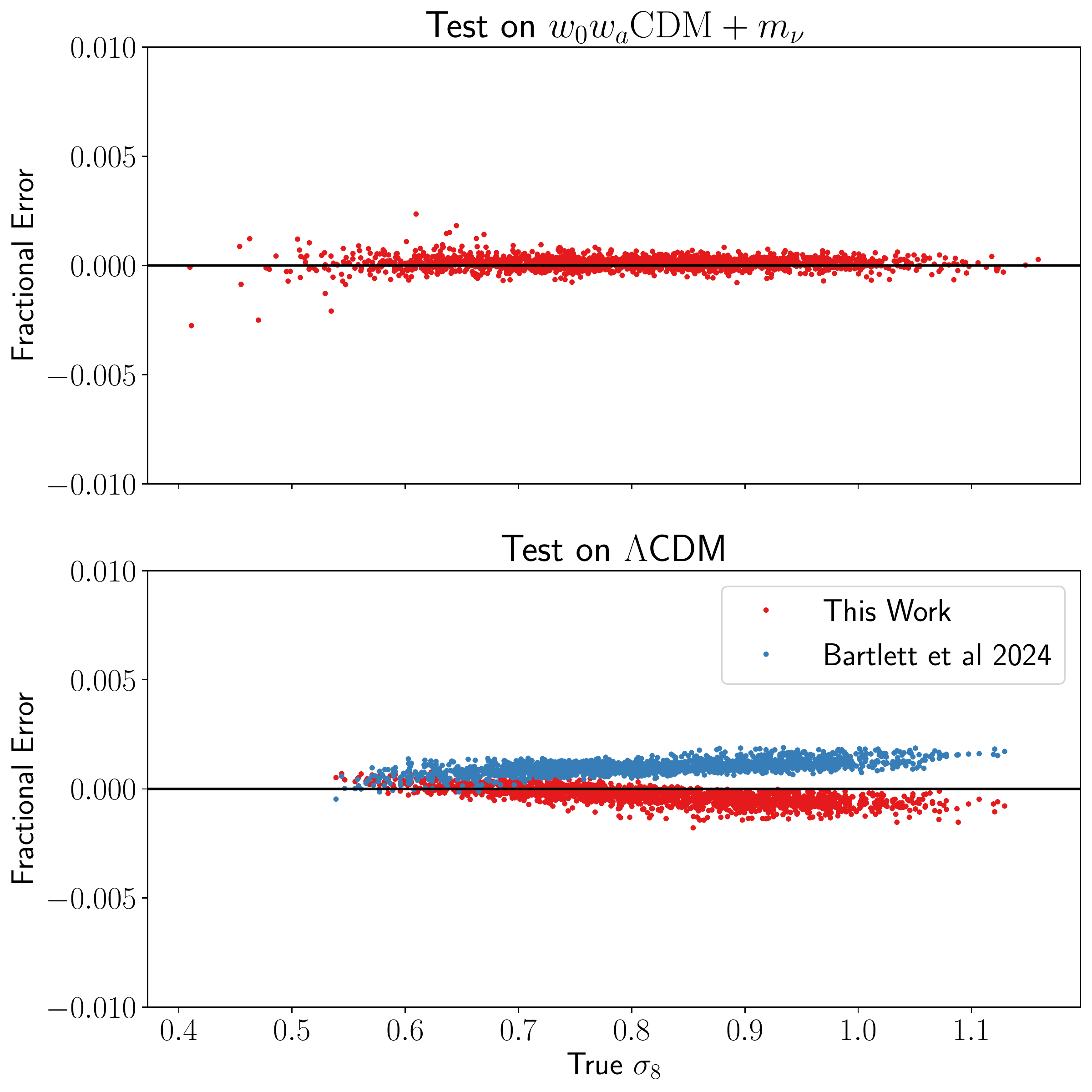}
    \caption{ Fractional error on the $\sigma_8$ prediction tested on $w_{0}w_{a}$CDM+$m_{\nu}$ (upper) and $\Lambda$CDM (bottom) models using the most accurate $\sigma_8$ expression (\cref{eq:sigma_8_accurate}). For the $\Lambda$CDM case, we also compare the results of the emulator given here to that of \citet{Bartlett_2024_linear}.}
    \label{fig:sigma8_max_precision}
\end{figure}

Here, we present an analytical approximation for $\sigma_{8}$ that, unlike \cref{eq:sigma8_result} which attempts to balance accuracy and simplicity, is the most accurate one found. This is the model at length 89 in \cref{fig:sigma8_pareto} and is given by
\afterpage{\FloatBarrier} 
\begin{widetext}
\begin{equation}
    \label{eq:sigma_8_accurate}
    \begin{split}
    \frac{\sigma_8}{\sqrt{10^9 A_{\rm s}}} & \approx 
     \tilde{c}_{0} 
     \left(\Omega_{m} \tilde{c}_{1} + \left( \tilde{c}_{2} m_{\nu} - \tilde{c}_{3} n_{s} + \log{\left(\tilde{c}_{4} h - \tilde{c}_{5} m_{\nu} \right)}\right) \left( \tilde{c}_{6} h + \tilde{c}_{7} m_{\nu} - \tilde{c}_{8} n_{s} + 1\right)\right) 
     \left(\tilde{c}_{9} h -  m_{\nu}\right)\\
     & \times \left( \tilde{c}_{10} w_{0} - \tilde{c}_{11} m_{\nu} - \left(\tilde{c}_{12} w_{0} - \tilde{c}_{13} w_{a} - \log{\left(\Omega_{m} \tilde{c}_{14} \right)}\right) \left(\Omega_{m} \tilde{c}_{15} + \tilde{c}_{16} w_{0} + \tilde{c}_{17} w_{a} + \log{\left(- \tilde{c}_{18} w_{0} - \tilde{c}_{19} w_{a} \right)}\right) \right. \\&
      \left. - \log{\left(\Omega_{m} \tilde{c}_{20} + \log{\left(- \tilde{c}_{21} w_{0} - \tilde{c}_{22} w_{a} \right)} \right)} + \log{\left(- \tilde{c}_{23} w_{0} - \tilde{c}_{24} w_{a} \right)}\right)\\
       & \times \left( \tilde{c}_{25} m_{\nu}- \sqrt{\Omega_{b}} \tilde{c}_{26} - \Omega_{b} \tilde{c}_{27} + \Omega_{m} \tilde{c}_{28} - \tilde{c}_{29} h  + 1 + \left( \Omega_{b} \tilde{c}_{30} - \tilde{c}_{31} h - \log{\left(\Omega_{m} \tilde{c}_{32} \right)}\right) \left(\Omega_{m} \tilde{c}_{33} - \tilde{c}_{34} h - \tilde{c}_{35} m_{\nu} - \tilde{c}_{36} n_{s}\right)\right),
    \end{split}
\end{equation}
\end{widetext}
\noindent 
where the parameters $\bm{\tilde{c}}$ are given in \cref{tab:sigma8_accruate_parameters}.
Although this form is rather long, we include it in case a more accurate conversion is required. We find that it has a root mean squared error on the $w_{0}w_{a}$CDM+$m_{\nu}$ and $\Lambda$CDM test sets of 0.02\% and 0.04\%, respectively. We plot the variation of this error with the true $\sigma_8$ value in \cref{fig:sigma8_max_precision}. For the $w_{0}w_{a}$CDM+$m_{\nu}$ model, the errors are significantly smaller compared to those in \cref{fig:sigma8_prediction}, while for $\Lambda$CDM, our results show comparable performance to \citet{Bartlett_2024_linear}.

\begin{table}[]
    \caption{Best-fit parameters for $\sigma_{8}$ given in \cref{eq:sigma_8_accurate}.}
    \centering
    \begin{tabular}{c|l|c|l|c|l}
    Param. & Value & Param. & Value & Param. & Value \\
    \hline\hline
    $\tilde{c}_0$  & 0.0246 & $\tilde{c}_{13}$ & 0.5466 & $\tilde{c}_{26}$ & 1.2991 \\
    $\tilde{c}_1$  & 2.1062 & $\tilde{c}_{14}$ & 0.5519 & $\tilde{c}_{27}$ & 4.1426 \\
    $\tilde{c}_2$  & 2.9355 & $\tilde{c}_{15}$ & 0.3689 & $\tilde{c}_{28}$ & 3.3055 \\
    $\tilde{c}_3$  & 0.7626 & $\tilde{c}_{16}$ & 0.3261 & $\tilde{c}_{29}$ & 0.5716 \\
    $\tilde{c}_4$  & 0.2962 & $\tilde{c}_{17}$ & 0.2002 & $\tilde{c}_{30}$ & 6.0094 \\
    $\tilde{c}_5$  & 0.5096 & $\tilde{c}_{18}$ & 0.8892 & $\tilde{c}_{31}$ & 1.9569 \\
    $\tilde{c}_6$  & 4.4025 & $\tilde{c}_{19}$ & 0.4462 & $\tilde{c}_{32}$ & 2.1477 \\
    $\tilde{c}_7$  & 3.6495 & $\tilde{c}_{20}$ & 1.215 & $\tilde{c}_{33}$ & 1.1902 \\
    $\tilde{c}_8$  & 0.4144 & $\tilde{c}_{21}$ & 3.4829 & $\tilde{c}_{34}$ & 0.128 \\
    $\tilde{c}_9$  & 0.8615 & $\tilde{c}_{22}$ & 2.5852 & $\tilde{c}_{35}$ & 0.6931 \\
    $\tilde{c}_{10}$ & 0.6188 & $\tilde{c}_{23}$ & 0.0242 & $\tilde{c}_{36}$ & 0.2661 \\
    $\tilde{c}_{11}$ & 0.1751 & $\tilde{c}_{24}$ & 0.0051 & & \\
    $\tilde{c}_{12}$ & 0.824 & $\tilde{c}_{25}$ & 0.1614 & & \\
    \end{tabular}
    \label{tab:sigma8_accruate_parameters}
\end{table}

\end{appendix}

\end{document}